\documentclass[aps,prx,twocolumn,superscriptaddress,showpacks]{revtex4-2}

\usepackage{comment}
\usepackage[colorlinks, citecolor=red]{hyperref}
\usepackage{hyperref}
\usepackage[utf8]{inputenc}
\usepackage[T1]{fontenc}    %
\usepackage[english]{babel}
\usepackage[pdftex]{graphicx}
\usepackage{amsmath,amssymb,amsthm,bbm, mathtools} %
\usepackage{mathrsfs}
\usepackage{tikz}   %
\usepackage[normalem]{ulem}

\renewcommand{\vec}[1]{\boldsymbol{#1}}

\begin{document}

\title{Theory of quantum decoherence in macroscopic topological insulators}

\author{Xian-Peng Zhang}

\affiliation{International Center for Quantum Materials, Beijing Institute of Technology, Zhuhai, 519000, China}

\affiliation{Centre for Quantum Physics, Key Laboratory of Advanced Optoelectronic Quantum Architecture and Measurement (MOE), School of Physics, Beijing Institute of Technology, Beijing, 100081, China}

\author{Yan-Qing Feng}
\email{yq_feng@bitzh.edu.cn}

\affiliation{International Center for Quantum Materials, Beijing Institute of Technology, Zhuhai, 519000, China}

%\author{Ji-Feng Shao}
%\email{shaojifeng@bitzh.edu.cn}

%\affiliation{International Center for Quantum Materials, Beijing Institute of Technology, Zhuhai, 519000, China}

%\author{Haiwen Liu}
%\affiliation{Center for Advanced Quantum Studies, School of Physics and Astronomy, Beijing Normal University, Beijing 100875, China}

\author{Wanxiang Feng}
%\email{wxfeng@bit.edu.cn}
\affiliation{Centre for Quantum Physics, Key Laboratory of Advanced Optoelectronic Quantum Architecture and Measurement (MOE), School of Physics, Beijing Institute of Technology, Beijing, 100081, China}
\affiliation{International Center for Quantum Materials, Beijing Institute of Technology, Zhuhai, 519000, China}

\author{Yugui Yao}
\email{ygyao@bit.edu.cn}
\affiliation{Centre for Quantum Physics, Key Laboratory of Advanced Optoelectronic Quantum Architecture and Measurement (MOE), School of Physics, Beijing Institute of Technology, Beijing, 100081, China}

\affiliation{International Center for Quantum Materials, Beijing Institute of Technology, Zhuhai, 519000, China}

\begin{abstract}
Quantum decoherence--the loss of quantum coherence due to interactions with an environment--plays a central role in quantum transport, and controlling this ubiquitous yet inevitable phenomenon is essential for practical quantum technologies. Despite its importance, the microscopic mechanisms of decoherence in \textit{infinite-size} topological insulators remain poorly understood. Here, we develop a comprehensive theory that quantitatively investigates how quantum decoherence shapes the quantum spin Hall effect in macroscopic topological insulators, and reveal that decoherence-induced corrections scale \textit{quadratically} with impurity density.  Besides, we uncover a previously unidentified mechanism of the extrinsic spin Hall effect: a \textit{second-order} skew-scattering process intrinsically tied to quantum decoherence--fundamentally distinct from, yet substantially stronger than, the conventional third-order skew-scattering mechanism. Furthermore, we predict a new scaling law in which the decoherence-induced spin Hall conductivity scales \textit{quadratically} with the longitudinal conductivity, providing a clear experimental signature of decoherence effects. Our results establish the essential role of  decoherence in quantum transport of topological insulators and reveal that macroscopic topological insulators offer a promising platform for next-generation spintronic applications. 
\end{abstract}

\maketitle

\textit{Introduction--}
Quantum coherence lies at the heart of quantum transport and underpins a wide range of emergent phenomena~\cite{von1986quantized,stormer1999nobel,yennie1987integral,laughlin1999nobel,chang2023colloquium,qi2010quantum}, including integer and fractional quantum Hall effects~\cite{von1986quantized,stormer1999nobel,yennie1987integral,laughlin1999nobel}, quantum anomalous Hall effect~\cite{chang2023colloquium,chang2013experimental}, and  quantum spin Hall effect (QSHE)~\cite{kane2005quantum,bernevig2006quantum,qian2014quantum,qian2014quantum,liu2008quantum,bernevig2006quantumexp,konig2007quantum,nowack2013imaging}. In quantum spin Hall insulators--for example, HgTe/CdTe~\cite{konig2007quantum,nowack2013imaging} or InAs/GaSb~\cite{liu2008quantum} quantum wells--helical edge states host counterpropagating electrons of opposite spin. Time-reversal symmetry forbids elastic backscattering from nonmagnetic disorder, so in principle helical edge channels can maintain long spin coherence time as long as phase coherence is preserved, leading to the precise quantization of transverse spin conductivity, regarded
as the smoking gun for the QSHE~\cite{konig2007quantum,bernevig2006quantum}. In practice, however, quantum coherence is inevitably degraded by spin-flip processes, electron–phonon scattering, and electron–electron interactions~\cite{chalker2007decoherence,jo2022scaling,nigg2016decoherence}. As phase coherence is lost, quantum superpositions can no longer interfere constructively, rendering edge transport dissipative and smearing the quantized conductance plateaus~\cite{yan2024rules,qi2019dephasing}. Concurrently, a finite longitudinal conductivity is almost invariably observed even when the Fermi energy lies deep within the bulk gap, often with strong temperature and disorder dependence~\cite{schmidt2013inelastic,helical2013vyrynen,vayrynen2014resistance}. This persistent bulk contribution obscures edge transport and limits the realization of ideal QSHE.

\begin{figure}[t]
\begin{center}
\includegraphics[width=0.48\textwidth]{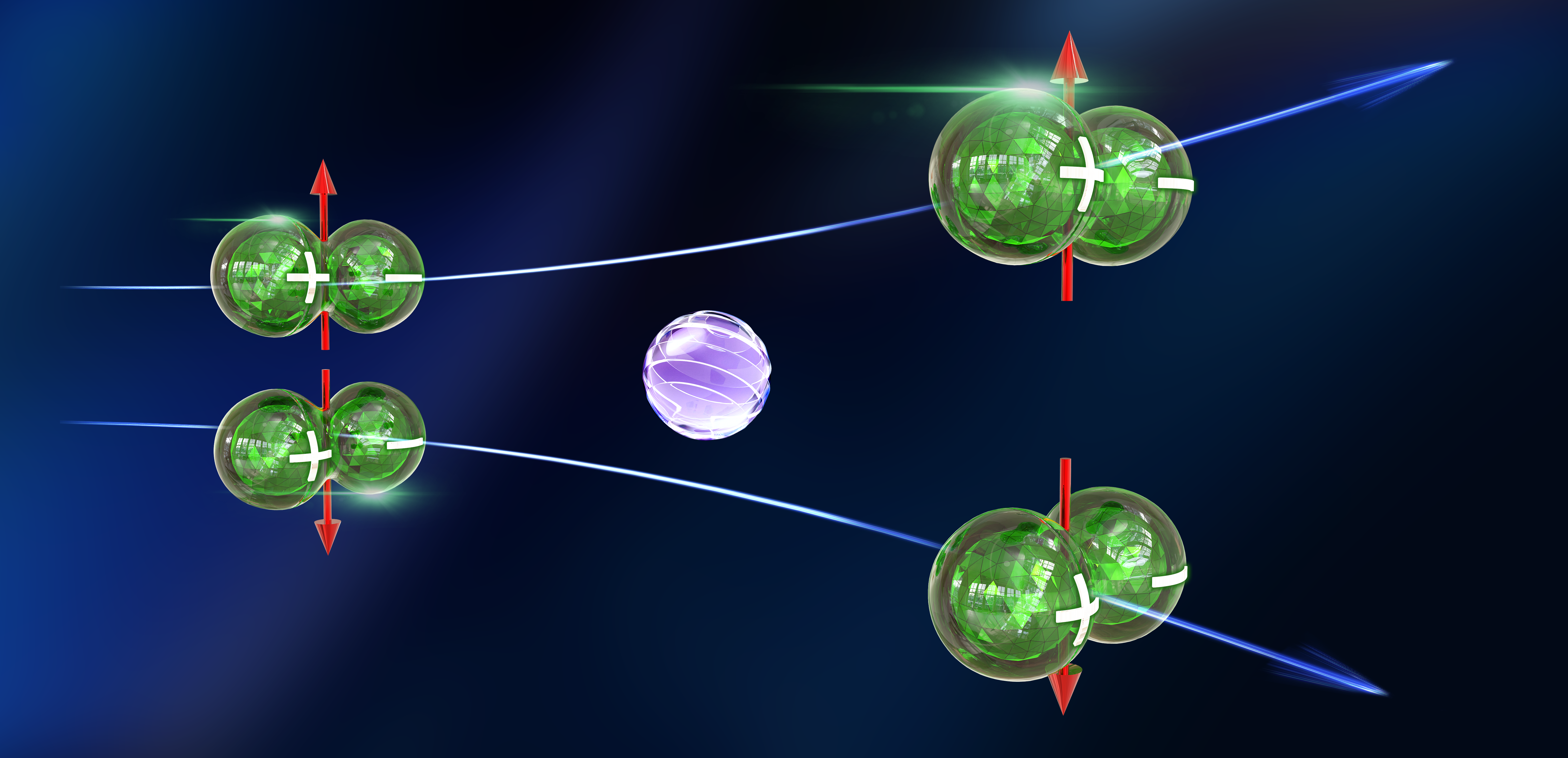} 
\end{center}
\caption{The sketch of the \textit{second-order} skew-scattering process intrinsically tied to quantum decoherence. }
\label{FIG1}
\end{figure}

Recent studies point to a microscopic origin of a longitudinal response rooted in quantum decoherence itself--the decay of the off-diagonal components of the density matrix~\cite{zhang2026magnetoresistance,zhang2026theory,zhang2026resistance}. In Berry-curvature-dominated systems, an applied electric field generically induces interband coherence across the entire Fermi sea via the Berry connection~\cite{zhang2026magnetoresistance,zhang2026theory,zhang2026resistance}, a direct consequence of spin–orbit coupling and band topology. In perfectly coherent limit, this coherence produces geometric velocities transverse to the applied field~\cite{xiao2010berry}, without contributing to longitudinal transport. The key insight is that disorder fundamentally alters this picture. Scattering processes that destroy phase coherence convert interband coherence into a finite longitudinal current~\cite{zhang2026magnetoresistance}, thereby opening an intrinsic bulk transport channel even in the absence of carriers at the Fermi energy. Physically, decoherence acts as a bridge between coherent interband dynamics and dissipative transport, enabling charge flow from states deep within the Fermi sea. This mechanism is distinct from conventional quasiparticle transport and does not rely on thermal activation. As a result, decoherence-induced bulk conduction can directly compete with--and mask--edge transport in topological insulators.

Here we develop a microscopic theory of quantum transport driven by decoherence in macroscopic topological insulators, where edge contributions are absent and the density of states vanishes at the Fermi energy. In this regime, conventional Drude transport is suppressed, leaving coherence-driven transport as the dominant mechanism. Using a quantum master-equation approach, we quantitatively establish how decoherence governs the QSHE, with particular emphasis on HgTe/CdTe quantum wells. Remarkably, we uncover a previously unrecognized mechanism of the extrinsic spin Hall effect (SHE): a second-order skew-scattering process intrinsically tied to quantum decoherence. This mechanism is fundamentally distinct from, and parametrically stronger than, the conventional third-order skew-scattering contribution. Our results provide a unified framework for understanding bulk transport in topological insulators and highlight decoherence not merely as a limiting factor, but as a fundamental driver of transport phenomena.

\textit{Model and theory--}
We investigate non-equilibrium quantum transport of itinerant electrons in topological insulators, described by the total Hamiltonian $\hat{H}=\hat{H}_{e}+\hat{H}_{E}+\hat{V}$~\cite{zhang2026resistance} with $\hat{H}_e=\sum_{ls} \hat{\mathcal{H}}^s_0(\vec{r}_{ls})$, $\hat{H}_{E}=-\sum_{ls} e\vec{E}\cdot \vec{r}_{ls}$, and $\hat{V}=\sum_{jls}U\delta(\vec{r}_{ls}-\vec{R}_j)$, where $e<0$ is the charge of electron, $\vec{r}_{ls}$ is the position operator of the $l$th electron with spin $s$, $\vec{R}_j$ is the position of the $j$th local moments that are randomly distributed and $U$ denotes the scattering strength of the impurities located at $\vec{R}_j$. The single-particle  Bernevig–Hughes–Zhang Hamiltonian of itinerant electrons is   $\hat{\mathcal{H}}^s_0(\vec{r}_{ls})=sAk_l^x\hat{\sigma}_l^x+Ak_l^y\hat{\sigma}_l^y+(M-Bk_l^2)\hat{\sigma}^z_l$~\cite{qi2010quantum,bernevig2006quantumexp}, where $\vec{\sigma}_l=(\hat{\sigma}_l^x,\hat{\sigma}_l^y,\hat{\sigma}_l^z)$ denotes Pauli matrices acting in orbital space.  Since the spin index $s=\uparrow,\downarrow$ is conserved, the problem decomposes into independent spin sectors. Diagonalization yields two isotropic bands $\epsilon_{\vec{k}\eta}=\eta\mathcal{E}_{\vec{k}}$, 
with $\mathcal{E}_{k}=\sqrt{A^2k^2+(M-Bk^2)^2}$. The corresponding eigenstates can be expressed in terms of angular parameters  $\vert \vec{k}s+ \rangle=\begin{bmatrix}
        s\cos\frac{\Theta_{k}}{2}e^{-si\theta_{\vec{k}}}&
        +\sin\frac{\Theta_{k}}{2}
    \end{bmatrix}^T$ and $\vert \vec{k}s- \rangle=\begin{bmatrix}
        s\sin\frac{\Theta_{k}}{2}e^{-si\theta_{\vec{k}}} &
        -\cos\frac{\Theta_{k}}{2}
    \end{bmatrix}^T$, 
with $\cos\Theta_{k}=(M-Bk^2)/\mathcal{E}_{k}$, $\sin\Theta_{k}=Ak/\mathcal{E}_{k}$, and $\theta_{\vec{k}}=\text{angle}\left(\frac{k_x+ik_y}{k}\right)$.  Here, we set $\vec{E}=E_{x}\vec{x}$ ($\vec{x}$ is unit vector in x direction). 

\begin{figure}[t]
\begin{center}
\includegraphics[width=0.48\textwidth]{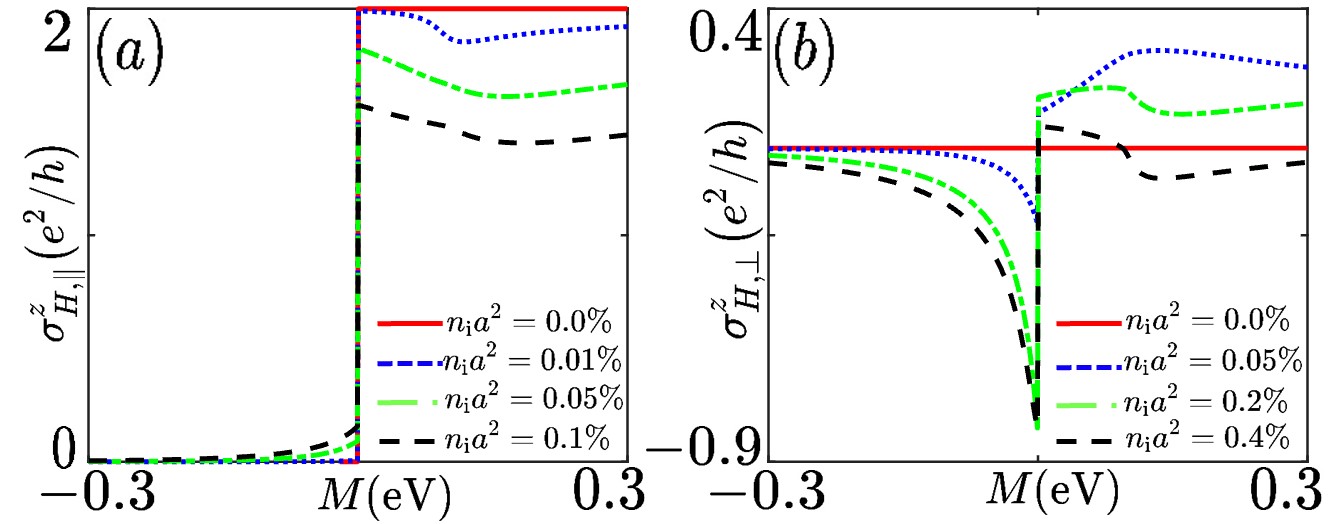} 
\end{center}
\caption{(a-b) The transverse spin conductivities derived from
(a) ordinary and (b) anomalous off-diagonal density matrix as a function of $M$ for several values of impurity density $n_{\text{i}}$. Here, $n_{\text{i}}$ is quantified by the energy unit $n_{\text{i}}U$. Other parameters are from the experimental results Ref.~\cite{bernevig2006quantumexp}: $A\pi/a=1.1$eV and  $B(\pi/a)^2=7.0$eV, $U/a^2=23$eV and $a=0.65$nm.}
\label{FIG}
\end{figure}

\textit{QSHE from perfect coherence--}
In a clean enough topological insulator, impurity scattering becomes ignorable. The off-diagonal component of density matrix, quantifying the quantum coherence between conduction and valence bands, is  proportional to the Berry connection ($\vec{\mathcal{R}}^{\bar{\eta}\eta}_{\vec{k}s}$) and, in steady state, is given by 
\begin{align} \label{agflalgt}
    \delta\varrho^{\bar{\eta}\eta}_{\vec{k}s}=-\frac{ f_{\vec{k}+}- f_{\vec{k}-}}{2\mathcal{E}_{\vec{k}}}\vec{\mathcal{R}}^{\bar{\eta}\eta}_{\vec{k}s}\cdot e\vec{E}.
\end{align}
The detailed expressions of the Berry connection $\vec{\mathcal{R}}^{\eta\eta'}_{\vec{k}s}=\langle \vec{k}s\eta\vert i\vec{\nabla}_{\vec{k}}\vert \vec{k}s\eta'\rangle$~\cite{xiao2010berry} are given in Appendix~\ref{berryconnectiofn}. Physically, when an electric field drives the system, electrons do not simply move within a single band; instead, their quantum states develop a coherent superposition across bands throughout the entire Fermi sea~\cite{culcer2017interband,sekine2017quantum,atencia2022semiclassical}. This electric field-induced quantum coherence survives within the \textit{whole} Fermi sea, producing a transverse flow of spin without any accompanying charge current  \footnote{The associated linear-in-$E$ conductivities  are calculated by using the equilibrium velocity $\hat{v}^{0}_{i,\vec{k}}$ (nonequilibrium coherence $\delta\varrho^{\bar{\eta}\eta}_{\vec{k}s}$) independent of (linear in) $\vec{E}$} 
\begin{align} \label{bgbgbsbz}
     \sigma^{z}_{H}=\frac{e^2}{h}\left\{\begin{matrix*}
       2, & \text{sign}(MB)>0\\
       0, & \text{sign}(MB)<0
    \end{matrix*}\right..
\end{align}
The resulting spin Hall response then reflects the intrinsic topology of the band structure, yielding the quantized value characteristic of the QSHE in the topologically nontrivial regime [$\text{sign}(MB)>0$], as illustrated by the red curve in Fig.~\ref{FIG}(a). From a physical perspective, this quantization arises because the electronic wave functions carry a nontrivial momentum-space spin texture. As electrons are accelerated by the electric field, their spin orientation evolves smoothly across momentum space, and this coherent spin dynamics generates a transverse spin current that is protected by quantum coherence.
However, experimental measurements typically deviate from perfect quantization due to unavoidable quantum decoherence, and the quantized conductance plateaus become smeared in the presence of environmental interactions in realistic systems~\cite{chalker2007decoherence,jo2022scaling,nigg2016decoherence}. The microscopic mechanisms of quantum decoherence in macroscopic topological insulators remain poorly understood, and their quantitative treatment is widely regarded as a formidable task. Consequently, a phenomenological scattering rate $\Gamma$, intended to parameterize quantum decoherence, is often introduced by hand~\cite{nagaosa2010anomalous,zhou2022transport,jiang2009topological,go2020orbital,czaja2014anomalous}
\begin{align}  \label{fvdfva}
    \sigma^{z}_{H,0}&=-\frac{e^2\hbar}{V} \sum_{\vec{k}s\eta\neq\eta'}s(f_{\vec{k}\eta}-f_{\vec{k}\eta'})\\
    &\times\text{Im}\left\{\frac{\langle \vec{k}s\eta\vert \hat{v}^x_{\vec{k}s}\vert \vec{k}s\eta'\rangle \langle \vec{k}s\eta'\vert  \hat{v}^y_{\vec{k}s}\vert \vec{k}s\eta\rangle}{(\epsilon_{\vec{k}\eta'}-\epsilon_{\vec{k}\eta}+i\Gamma)^2}\right\}, \notag
\end{align}
which reduces to Eq.~\eqref{bgbgbsbz} when $\Gamma\to0$. However, this phenomenological Kubo formula does not capture the momentum dependence of the decoherence rate $\Gamma$, thereby preventing a quantitative description of the QSHE.

\textit{Theory of quantum decoherence--} To incorporate the quantum decoherence -- the relaxation of the off-diagonal density matrix, we here go beyond the previous results of Ref.~\cite{culcer2017interband, culcer2022anomalous}, i.e.,  Eq.~\eqref{agflalgt} and assume the following ansatz for off-diagonal density matrix into the collision integral, which 
are described by two unknown complex parameters $\tau^{\bar{\eta}\eta}_{ks,\Vert}$ and $\tau^{\bar{\eta}\eta}_{ks,\perp}$, dubbed as ordinary and anomalous decoherence time~\cite{zhang2026theory,zhang2026magnetoresistance}
\begin{align} \label{ansatz}
    \delta\varrho^{\bar{\eta}\eta}_{\vec{k}s}=\delta\varrho^{\bar{\eta}\eta}_{\vec{k}s,\Vert}+\delta\varrho^{\bar{\eta}\eta}_{\vec{k}s,\perp},
\end{align}
with
\begin{align} \label{fvfkvmkdf4}
\delta\varrho^{\bar{\eta}\eta}_{\vec{k}s,\Vert/\perp}=-\frac{e}{\hbar}\tau^{\bar{\eta}\eta}_{ks,\Vert/\perp}(f_{k\bar{\eta}}- f_{k\eta})\vec{\mathcal{R}}^{\bar{\eta}\eta}_{\vec{k}s}\cdot \vec{E}_{\Vert/\perp},
\end{align}
where $\vec{E}_{\Vert}=\vec{E}$ and $\vec{E}_{\perp}=\vec{E}\times \hat{z}$. Following the methodology of Refs.~\cite{zhang2024microscopic,zhang2026magnetoresistance,zhang2026theory,zhang2025open}, we derive the collision integral that contains the off-diagonal density matrix with anzatz \eqref{ansatz}, and the off-diagonal component of the quantum kinetic equation becomes (see detailed derivations in Appendix~\ref{qke} and Appendix~\ref{dkfvkdk})
\begin{align} \label{vfkfvkmd}
    \frac{\partial}{\partial t}  \delta\varrho^{\bar{\eta}\eta}_{\vec{k}s}&+\frac{i}{\hbar}(\epsilon_{k\bar{\eta}}-\epsilon_{k\eta}) \delta\varrho^{\bar{\eta}\eta}_{\vec{k}s}+e\frac{i}{\hbar}E^i\mathcal{R}^{i,\bar{\eta}\eta}_{\vec{k}s} (f_{k\bar{\eta}}- f_{k\eta}) \notag \\
    &=-\frac{1}{\hbar}\Gamma_{k} \delta\varrho^{\bar{\eta}\eta}_{\vec{k}s}+is\eta\frac{1}{\hbar}\Gamma^a_{k}\delta\varrho^{\bar{\eta}\eta}_{\vec{k}s,a},
\end{align}
with
\begin{align} \label{gbbkgbm}
    \delta\varrho^{\bar{\eta}\eta}_{\vec{k}s,a}=\frac{\tau^{\bar{\eta}\eta}_{ks,\Vert}}{\tau^{\bar{\eta}\eta}_{ks,\perp}}\delta\varrho^{\bar{\eta}\eta}_{\vec{k}s,\perp}-\frac{\tau^{\bar{\eta}\eta}_{ks,\perp}}{\tau^{\bar{\eta}\eta}_{k,\Vert}}\delta\varrho^{\bar{\eta}\eta}_{\vec{k}s,\Vert}.
\end{align}
The imaginary, antisymmetric components--the second term on the right-hand side of Eq.~\eqref{vfkfvkmd}--acquire opposite signs for opposite spins. For electrons in coherent superposition states, this anomalous scattering mechanism can be viewed intuitively as an effective out-of-plane magnetic field generated during the scattering event. This emergent field acts with opposite signs on opposite spins, thereby pushing them toward opposite transverse directions. Consequently, the scattering probabilities become asymmetric: spin-up electrons tend to scatter upward, while spin-down electrons preferentially scatter downward, as illustrated in Fig.~\ref{FIG1}. Importantly, this process represents a form of second-order skew-scattering that relies on interband coherence. It is therefore fundamentally different from, yet substantially stronger than, the conventional skew-scattering mechanism, which refers to quantum
distribution–the diagonal component of the density matrix,  appearing only at third order in impurity scattering and typically producing a much weaker response.  
The ordinary scattering causes quantum decoherence, i.e., the decay of the off-diagonal density matrix, which is quantified by a normal decoherence rate $\Gamma_{k}=\frac{\hbar}{4\tau^{0}_{k}}\left[1+\frac{(M-Bk^2)^2}{\mathcal{E}^2_{k}}\right]$, where $1/\tau^0_{k}=(2\pi/\hbar)n_{\text{i}}\nu_{k}U^{2}$. The density of states is given by $\nu_{k}=\frac{\mathcal{E}_k}{2\pi\vert A^2-2B(M-Bk^2)\vert }$. More importantly, the anomalous skew-scattering, quantified by an anomalous decoherence rate $\Gamma^a_{k}=\frac{\hbar}{2\tau^{0}_{k}} \frac{(M-Bk^2)}{\mathcal{E}_k}$, generates an effective out-of-plane magnetic field that points in opposite directions for different spins [the last term of the right-hand side of Eq.~\eqref{vfkfvkmd}]. 
By substitution of Eqs.~\eqref{ansatz}, ~\eqref{fvfkvmkdf4}, and \eqref{gbbkgbm}, equation~\eqref{vfkfvkmd}, in steady state, is separated into ordinary and anomalous components of density matrix, i.e., $\delta\varrho^{\bar{\eta}\eta}_{\vec{k}s,\Vert}$ and $\delta\varrho^{\bar{\eta}\eta}_{\vec{k}s,\perp}$, and we attain \textit{spin-dependent} normal and anomalous decoherence time as follows~\cite{zhang2026magnetoresistance,zhang2026theory}  
\begin{align} \label{sgfbf1}
    \frac{\tau^{\bar{\eta}\eta}_{ks,\Vert}}{\hbar}=\frac{(\epsilon_{k\bar{\eta}}-\epsilon_{k\eta})-i\Gamma_{k}}{[(\epsilon_{k\bar{\eta}}-\epsilon_{k\eta})-i\Gamma_{k}]^2+(\Gamma^a_{k})^2},
\end{align}
\begin{align} \label{sgfbf2}
    \frac{\tau^{\bar{\eta}\eta}_{ks,\perp}}{\hbar}=\frac{s\eta\Gamma^a_{k}}{[(\epsilon_{k\bar{\eta}}-\epsilon_{k\eta})-i\Gamma_{k}]^2+(\Gamma^a_{k})^2}.
\end{align}
From Eq.~\eqref{sgfbf1} and Eq.~\eqref{sgfbf2}, we see the normal decoherence rate $\Gamma_{k}$,  traditionally regarded as phenomenological parameter in the Kubo formula~\eqref{fvdfva}, describes the bandwidth of both ordinary and anomalous off-diagonal density matrix, that are coupled with each other via anomalous decoherence rate $\Gamma^a_{k}$. Therefore, both $\Gamma_{k}$ and $\Gamma^a_{k}$ play an important role in the quantum transport associated quantum coherence, by which we can analyze and effectively tune the ubiquitous quantum decoherence in topological materials. Notably, the electric field-induced quantum coherence [Eqs,~\eqref{ansatz} and \eqref{fvfkvmkdf4}] survives within the \textit{whole} Fermi sea, even though the density of states vanishes at the Fermi energy,  contributing a longitudinal charge current~\cite{zhang2026magnetoresistance,zhang2026theory,zhang2026resistance} and a transverse spin current in the presence of the impurity-induced collisional dissipation. We have demonstrated a decoherence-induced conductivity that scales \textit{linearly} with impurity density $n_{\text{i}}$~\cite{zhang2026resistance}, in stark contrast to the conventional Drude contribution, which is inversely proportional to $n_{\text{i}}$. This scaling implies that even a small concentration of impurities can generate an appreciable longitudinal conductivity, providing a natural explanation for the experimentally observed finite longitudinal response even when the Fermi energy lies within the bulk gap, particularly at low temperatures~\cite{konig2007quantum,knez2010finite,shamim2021quantized}. Here, we extend this framework to investigate the role of decoherence in the QSHE.

\begin{figure}[t]
\begin{center}
\includegraphics[width=0.48\textwidth]{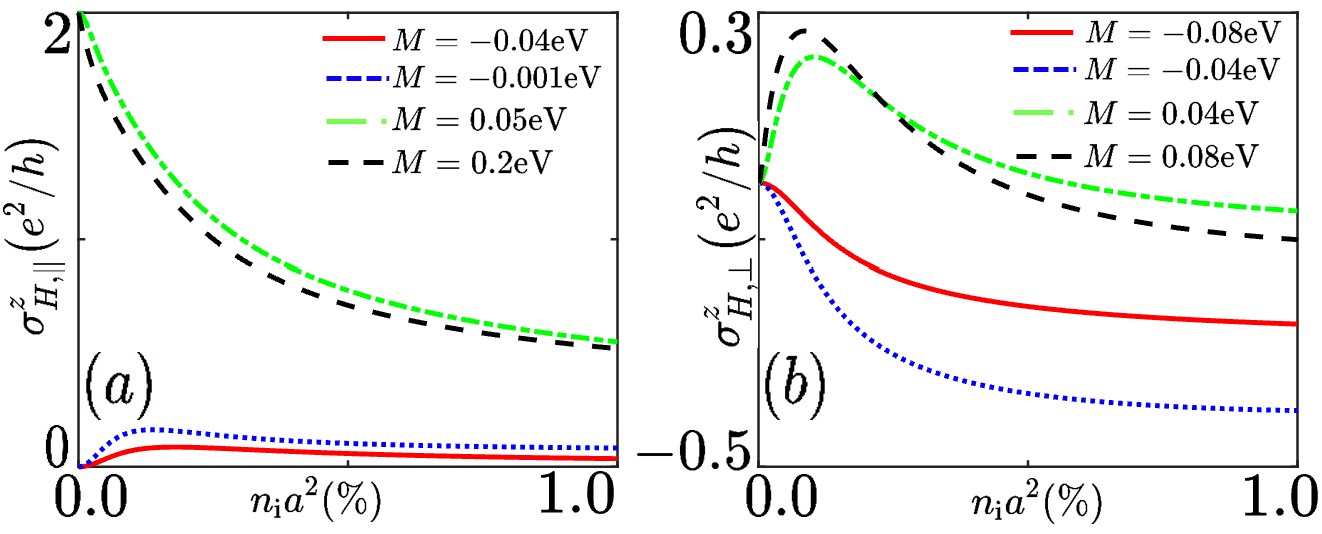} 
\end{center}
\caption{(a-b) The transverse spin conductivities derived from
(a) ordinary and (b) anomalous off-diagonal density matrix as a function of impurity density $n_{\text{i}}$ for several values of $M$.  Other parameters are the same as Fig. \ref{FIG}.}
\label{FIGSM1}
\end{figure}

\textit{QSHE in the presence of quantum decoherence--} We now examine how quantum decoherence modifies the QSHE. In general, the total spin Hall conductivity contains both ordinary and anomalous contributions, i.e., $\sigma^{z}_{H}=\sigma^{z}_{H,\Vert}+\sigma^{z}_{H,\perp}$. 
We first consider the ordinary off-diagonal density matrix $\delta\varrho^{\bar{\eta}\eta}_{\vec{k}s,\Vert}$, which already extends beyond the result in Eq.~\eqref{agflalgt} and generates a pure transverse spin current characterized by the spin Hall conductivity (see detailed derivations in Appendix~\ref{derivationcurrent})
\begin{align}  \label{zHov}
    \sigma^{z}_{H,\Vert}=+\frac{2e^2}{h}\int^{\infty}_{0}kdk\text{Re}\left\{\frac{\tau^{+-}_{k,\Vert}}{\hbar}\right\}\left[\frac{A^2(M+Bk^2)}{\mathcal{E}^2_k}\right],
\end{align}
where $\tau^{\bar{\eta}\eta}_{k,\Vert}=\tau^{\bar{\eta}\eta}_{ks,\Vert}$, which, in the absence of collision terms, reduces to the perfectly coherent limit [Eq.~\eqref{bgbgbsbz}], where the quantized spin Hall conductivity remains robust in the topologically nontrivial regime ($M>0$), as ensured by quantum coherence [red curve in Fig.~\ref{FIG}(a)]. However, impurity scattering introduces decoherence  gradually disrupting this coherent interband motion. Each scattering event acts as a random perturbation that partially erases the phase relation between the two bands.  In the dilute-impurity limit, the real part of the normal decoherence time becomes $\text{Re}(\tau^{+-}_{k,\Vert}/\hbar)\simeq 1/(2\mathcal{E}_k)-[\Gamma^2_k+(\Gamma^a_k)^2]/(8\mathcal{E}^3_k)$. Since both $\Gamma_k^a\propto n_{\text{i}}$ and $\Gamma_k\propto n_{\text{i}}$, the decoherence-induced correction to the QSHE, defined by $\delta \sigma^{z}_{H,\Vert}=\sigma^{z}_{H,\Vert}-\sigma^{z}_{H,0}$, grows only quadratically with impurity density $n_{\text{i}}$  in either the topologically nontrivial or trivial regimes [Fig.~\ref{FIGSM1}(a)]. This quadratic dependence implies that weak disorder has only a minor impact on the QSHE: the coherent spin transport remains largely intact unless impurities become sufficiently abundant. Consequently, a sizable impurity concentration is required before quantum decoherence produces an observable deviation from the ideal quantized response.

Quantitatively, Figure~\ref{FIG}(a) plots $\sigma_{H,\Vert}^{z}$ for different levels of impurity-induced dissipation. The case $n_{\text{i}}a^2=0.1\%$ corresponds to $n_{\text{i}}U=12$ meV, placing the system within the weak-disorder regime of our model. This regime represents a situation where impurities are sparse enough that electrons still propagate coherently over relatively long distances, but scattering events begin to introduce measurable decoherence. A contrasting behavior emerges in the topologically nontrivial regime. There, the spin Hall conductivity becomes noticeably more sensitive to decoherence. From a physical perspective, the intrinsic spin Hall response in the topological phase originates from the band structure itself—specifically from the momentum-space structure of the electronic states. Impurity scattering disrupts the coherent motion of these states, thereby weakening the intrinsic mechanism more effectively than in the trivial phase. This behavior follows from the integrand of Eq.~\eqref{zHov}, which is proportional to $(M+Bk^2)$ and therefore becomes larger when $M$ and $B$ share the same sign. Consequently, the spin Hall signal is suppressed more rapidly by decoherence when the system is topologically nontrivial. This trend differs from earlier phenomenological numerical studies of mesoscopic topological insulators~\cite{jiang2009topological,Yanxia2008}, which suggested a different robustness of the topological response. Our results therefore highlight that decoherence plays a decisive role in determining the observable quantum spin Hall response, reflecting the competition between impurity-driven asymmetric scattering and topology-driven intrinsic transport.

\textit{Extrinsic SHE from second-order skew scattering--}We next examine the contribution arising from the anomalous off-diagonal density matrix $\delta\varrho^{\bar{\eta}\eta}_{\vec{k}s,\perp}$. The corresponding transverse spin Hall conductivity is given by (see detailed derivations in Appendix~\ref{derivationcurrent})
\begin{align} \label{zHop}
   \sigma^{z}_{H,\perp}=-\frac{e^2}{h}\int^{\infty}_{0}kdk\text{Im}\left\{\frac{\tau^{+-}_{k,\perp}}{\hbar}\right\}\left[\frac{A^2(M+Bk^2)^2}{\mathcal{E}^3_k}+\frac{A^2}{\mathcal{E}_k}\right],
\end{align}
where $\tau^{+-}_{k,\perp}=\tau^{+-}_{k\uparrow,\perp}$. This term originates from impurity-induced asymmetric scattering acting on quantum-coherent electronic states. When electron simultaneously occupies conduction and valence bands, electrons are slightly more likely to be deflected to one transverse side depending on their spin orientation (Fig. \ref{FIG1}). Although the deflection from a single impurity is small, the cumulative imbalance from many scattering events produces a transverse spin current. The conductivity $\sigma^{z}_{H,\perp}$ is governed by the imaginary part of the anomalous decoherence time [Eq.~\eqref{sgfbf2}]. In the dilute-impurity limit, $\text{Im}(\tau^{+-}_{k,\perp})$ is proportional to both $\Gamma_k$ and $\Gamma_k^a$, i.e., $\text{Im}(\tau^{+-}_{k,\perp})\propto \Gamma_k^a\Gamma_k$. Consequently, the decoherence-induced contribution $\sigma^{z}_{H,\perp}$ grows quadratically with impurity density ($\sigma^{z}_{H,\perp}\propto n_\text{i}^2$) in both the topologically nontrivial and trivial regimes [Fig.~\ref{FIGSM1}(b)]. 

The quantitative behavior of $\sigma^{z}_{H,\perp}$ is shown in Fig.~\ref{FIG}(b). Because $\text{Im}(\tau^{+-}_{k,\perp})\propto\Gamma_k^a$, the magnitude of $\sigma^{z}_{H,\perp}$ depends sensitively on the anomalous decoherence rate $\Gamma_k^a$. In the absence of impurity scattering, the asymmetric scattering channel is absent, and therefore $\sigma^{z}_{H,\perp}$ vanishes [red curve in Fig.~\ref{FIG}(b)]. As the impurity density increases, asymmetric scattering becomes more prominent and generates a sizable transverse spin current. Numerically, $\sigma^{z}_{H,\perp}$ can reach values as large as $\sim -0.9 e^2/h$ near $M\to 0_{-}$ for $n_\text{i} a^2=0.4\%$ [black curve in Fig.~\ref{FIG}(b)], while a substantial contribution $\sim0.3e^2/h$ persists in the $M>0$ regime. Notably, the corrections from $\delta\sigma^{z}_{H,\Vert}$ and $\sigma^{z}_{H,\perp}$ may carry opposite signs, leading to a partial cancellation of the overall decoherence-induced modification to the spin Hall response, as illustrated in Fig.~\ref{FIGSM1}. This observation highlights two important limitations of the phenomenological Kubo approach [Eq.~\eqref{fvdfva}] and/or Green function treatment. First, it neglects the strong momentum dependence of the decoherence time. Second, it fails to capture the compensating contribution from $\sigma^{z}_{H,\perp}$ and therefore tends to overestimate the impact of decoherence (for instance, $\sigma^{z}_{H,\perp}\sim -0.9e^2/h$ at $M=0^{-}$). Our formalism thus provides a more accurate framework for quantitatively describing how impurity scattering and quantum coherence together shape the spin Hall response.

\textit{Scaling analysis--}Finally, we examine how decoherence effects can be distinguished from intrinsic contributions. While $\sigma^{z}_{H,\Vert}$ does not exhibit a simple scaling behavior, the decoherence-induced correction to the QSHE, $\delta \sigma^{z}_{H,\Vert}$, scales quadratically with impurity density $n_{\mathrm{i}}$ in the dilute limit, as $\sigma^{z}_{H,\perp}$ and the total correction $\delta \sigma^z_{H}=\delta \sigma^{z}_{H,\Vert}+\sigma^{z}_{H,\perp}$ do. This distinct dependence on impurity concentration provides an experimentally accessible criterion to separate our mechanism from intrinsic and side-jump contributions, which remain largely insensitive to $n_{\mathrm{i}}$~\cite{sinova2015spin,berger1970side,berger1972application,lyo1972side}, as well as from skew-scattering processes that instead scale as $1/n_{\mathrm{i}}$~\cite{engel2005theory,niimi2012giant,ferreira2014extrinsic}. In macroscopic topological insulators, the density of states vanishes at the Fermi level, suppressing the Drude contribution. In this regime, the longitudinal response induced by decoherence follows $\sigma_{L}\propto n_{\mathrm{i}}$~\cite{zhang2026resistance,zhang2026magnetoresistance,zhang2026theory}, in stark contrast to the conventional Drude behavior $\sigma_L \propto 1/n_{\mathrm{i}}$. These relations lead to a scaling law $\delta \sigma^z_{H} \propto \sigma_L^2$, providing a clear experimental signature of decoherence effects.

\textit{Conclusion--}We have developed a microscopic theory of quantum transport driven by decoherence in macroscopic topological insulators, where edge contributions are absent and the density of states vanishes at the Fermi energy. Within this framework, we quantitatively establish how quantum decoherence governs the QSHE. We further identify a previously unrecognized mechanism for the extrinsic SHE, arising from a second-order skew-scattering process intrinsically linked to decoherence-fundamentally distinct from, and parametrically stronger than, the conventional third-order mechanism. Our results highlight quantum decoherence not merely as a limiting factor, but as a fundamental driver of transport. Controlling decoherence is therefore essential, both for realizing robust quantum spin Hall transport in next-generation spintronic devices and for advancing fault-tolerant quantum computation based on topologically protected states.

\textit{Acknowledgements--}
This work is supported by National Key R$\&$D Program of China (Grant Nos. 2020YFA0308800, 2021YFA1401500), the National Natural Science Foundation of China (Grant Nos. 12234003, 12321004, 12022416, 12475015, 11875108), the National Council for Scientific and Technological Development (Grant No. 301595/2022-4), and the Guangdong Basic and Applied Basic Research Foundation (Grants No. 2024A1515030118).

\appendix

\section{Quantum kinetic equation} \label{qke}

Here, we study quantum transport arising from the electric field-induced coherence in macroscopic topological insulators. Thus, we focus on the time evolution of the off-diagonal component of the non-equilibrium  density matrix~\cite{culcer2017interband,atencia2022semiclassical,zhang2026theory}
\begin{align} \label{fvdvkfvkmain1}
    \frac{\partial}{\partial t}  \delta\varrho^{\bar{\eta}\eta}_{\vec{k}s}-\eta\frac{2i\mathcal{E}_{\vec{k}}}{\hbar} \delta\varrho^{\bar{\eta}\eta}_{\vec{k}s}+\frac{i}{\hbar}\vec{\mathcal{R}}^{\bar{\eta}\eta}_{\vec{k}s} (f_{k\bar{\eta}}- f_{k\eta})\cdot e\vec{E}= \mathcal{ J}^{\bar{\eta}\eta}_{\vec{k}s}(\delta\varrho),
\end{align}
where $\hbar$ is the reduced Plank constant. The second term of the left hand side of Eq.~\eqref{fvdvkfvkmain1} is dynamic term,  while the third one is drive term. The Berry connections, defined by $\vec{\mathcal{R}}^{\bar{\eta}\eta}_{\vec{k}s}=i\langle \vec{k}s\bar{\eta}\vert \partial_{\vec{k}}\vert \vec{k}s\eta\rangle $, reads(see derivations in Appendix.~\ref{berryconnectiofn}) 
\begin{align}
\mathcal{R}^{\bar{\eta}\eta}_{x,\vec{k}s}=-s\frac{A}{2\mathcal{E}_{k}}\sin \theta_{\vec{k}}-i\eta\frac{AE^+_{k}}{2\mathcal{E}^2_k}\cos\theta_{\vec{k}},
\end{align}

\begin{align}
\mathcal{R}^{\bar{\eta}\eta}_{y,\vec{k}s}=+\frac{A}{2\mathcal{E}_{k}}\cos \theta_{\vec{k}}-i\eta\frac{AE^+_{k}}{2\mathcal{E}^2_k}\sin\theta_{\vec{k}},
\end{align}
The collision term, within second-order \textit{Born-Markov approximation}, is given by~\cite{zhang2024microscopic,zhang2025open} 
\begin{align} \label{fdfkvIavf}
    \mathcal{ J}^{\bar{\eta}\eta}_{\vec{k}s}(\delta\varrho)&= \frac{\pi n_{\text{i}}U^2}{\hbar^2V}\sum_{\vec{k}'} \left\lbrace \left[\delta(\omega^{\vec{k}\eta}_{\vec{k}'\eta}) \sigma^s_{\vec{k}\bar{\eta},\vec{k}'\eta''} \delta\varrho_{\vec{k}'s}^{\bar{\eta}\eta} \sigma^s_{\vec{k}'\eta',\vec{k}\eta}  \right. \right. \notag \\
    &+\left.\delta(\omega^{\vec{k}'\bar{\eta}}_{\vec{k}\bar{\eta}})  \sigma^s_{\vec{k}\bar{\eta},\vec{k}'\bar{\eta}}  \delta\varrho_{\vec{k}'s}^{\bar{\eta}\eta}   \sigma^s_{\vec{k}'\eta,\vec{k}\eta} \right] \notag \\
  &-\left[\delta(\omega^{\vec{k}\bar{\eta}}_{\vec{k}'\bar{\eta}})  \sigma^s_{\vec{k}\bar{\eta},\vec{k}'\bar{\eta}}  \sigma^s_{\vec{k}'\bar{\eta},\vec{k}\bar{\eta}}    
   \delta\varrho_{\vec{k}s}^{\bar{\eta}\eta}\right. \notag\\
  &\left.\left.+ \delta(\omega^{\vec{k}'\eta}_{\vec{k}\eta}) \delta\varrho_{\vec{k}s}^{\bar{\eta}\eta} \sigma^s_{\vec{k}\eta,\vec{k}'\eta}   \sigma^s_{\vec{k}'\eta,\vec{k}\eta}\right]\right\}.  
\end{align}
with $\sigma^s_{\vec{k}'\eta',\vec{k}\eta}=\langle \vec{k}'s\eta'\vert \sigma^o\vert \vec{k}s\eta \rangle $. Here, we include only intraband scattering processes, since direct interband scattering is energetically forbidden in the weak-disorder limit of our model.   Both the electric field and scattering potential induce band-mixing effects, i.e., the quantum coherence. The former cause electric field-induced decoherence, while the later dissipates the gain coherence,  which are reflected in drive term and collision term, respectively. At steady state, the electric field-induced coherence gain is exactly balanced by impurity-induced collisional dissipation.

\section{Derivations of the Berry connection, Berry curvature  and velocity operators} \label{berryconnectiofn}

\subsection{Derivations of the Berry connection}

Here, we define the following spin Berry connection in
the $k$-space 
\begin{align} \label{berryconnection}
    \mathcal{R}^{\eta_i\eta_f}_{\vec{k}s}=\langle \vec{k}s\eta_i\vert (i\partial_{\vec{k}}\vert \vec{k}s\eta_f\rangle).
\end{align}
By substitution of eigenstates, i.e.,
\begin{align} \label{fdvefk1}
    \vert \vec{k}s+ \rangle=\begin{bmatrix}
        s\cos\frac{\Theta_{k}}{2}e^{-si\theta_{\vec{k}}}\\
        +\sin\frac{\Theta_{k}}{2}
    \end{bmatrix},
\end{align}
\begin{align} \label{fdvefk2}
    \vert \vec{k}s- \rangle=\begin{bmatrix}
        s\sin\frac{\Theta_{k}}{2}e^{-si\theta_{\vec{k}}}\\
        -\cos\frac{\Theta_{k}}{2}
    \end{bmatrix},
\end{align}
the matrix elements of the Berry connection \eqref{berryconnection} become
\begin{align} 
    \mathcal{R}^{++}_{i,\vec{k}s}=\frac{i}{2}(1+\cos\Theta_{k})e^{+si\theta_{\vec{k}}}\partial_{k_i}e^{-si\theta_{\vec{k}}},
\end{align}
\begin{align} 
    \mathcal{R}^{+-}_{i,\vec{k}s}
    =\frac{i}{2}\sin\Theta_{k}e^{+si\theta_{\vec{k}}}\partial_{k_i}e^{-si\theta_{\vec{k}}}+\frac{i}{2}\partial_{k_i}\Theta_{k},
\end{align}
\begin{align} 
    \mathcal{R}^{-+}_{i,\vec{k}s}
    =\frac{i}{2}\sin\Theta_{k}e^{+si\theta_{\vec{k}}}\partial_{k_i}e^{-si\theta_{\vec{k}}}-\frac{i}{2}\partial_{k_i}\Theta_{k},
\end{align}
\begin{align}
    \mathcal{R}^{--}_{i,\vec{k}s}=\frac{i}{2}(1-\cos\Theta_{k})e^{+si\theta_{\vec{k}}}\partial_{k_i}e^{-si\theta_{\vec{k}}}.
\end{align}
During the above derivations,  we have used the following identities
\begin{align}
+\frac{1}{2}\partial_{k_x}\Theta_{k}
&=\cos\frac{\Theta_{k}}{2}\partial_{k_x}\sin\frac{\Theta_{k}}{2}-\sin\frac{\Theta_{k}}{2}\partial_{k_x}\cos\frac{\Theta_{k}}{2},
\end{align}
\begin{align}
+\frac{1}{2}\partial_{k_y}\Theta_{k}
&=\cos\frac{\Theta_{k}}{2}\partial_{k_y}\sin\frac{\Theta_{k}}{2}-\sin\frac{\Theta_{k}}{2}\partial_{k_y}\cos\frac{\Theta_{k}}{2}.
\end{align}
Next, we calculate the $\frac{1}{2}\partial_{k_i}\Theta_{\vec{k}}$.  By means of the following four identities 
\begin{align}
    \partial_{k_x}\cos\frac{\Theta_{k}}{2}=-\frac{A^2}{4}\frac{1}{\cos\frac{\Theta_{k}}{2}}\frac{2Bk^2k_x+E_{k}k_x}{\mathcal{E}^3_k},
\end{align}
\begin{align}
    \partial_{k_y}\cos\frac{\Theta_{k}}{2}=-\frac{A^2}{4}\frac{1}{\cos\frac{\Theta_{k}}{2}}\frac{2Bk^2k_y+E_{k}k_y}{\mathcal{E}^3_k},
\end{align}
\begin{align}
    \partial_{k_x}\sin\frac{\Theta_{k}}{2}=+\frac{A^2}{4}\frac{1}{\sin\frac{\Theta_{k}}{2}}\frac{2Bk^2k_x+E_{k}k_x}{\mathcal{E}^3_k},
\end{align}
\begin{align}
    \partial_{k_y}\sin\frac{\Theta_{k}}{2}=+\frac{A^2}{4}\frac{1}{\sin\frac{\Theta_{k}}{2}}\frac{2Bk^2k_y+E_{k}k_y}{\mathcal{E}^3_k},
\end{align}
we attain
\begin{align}
+\frac{1}{2}\partial_{k_x}\Theta_{k}
&=\frac{A}{2k}\frac{2Bk^2k_x+E_{k}k_x}{\mathcal{E}^2_k}=\frac{A(M+Bk^2)}{2\mathcal{E}^2_k}\cos\theta_{\vec{k}},
\end{align}
\begin{align}
+\frac{1}{2}\partial_{k_y}\Theta_{k}
&=\frac{A}{2k}\frac{2Bk^2k_y+E_{k}k_y}{\mathcal{E}^2_k}=\frac{A(M+Bk^2)}{2\mathcal{E}^2_k}\sin\theta_{\vec{k}},
\end{align}
 By means of the following two identities
\begin{align}
    e^{+si\theta_{\vec{k}}}\partial_{k_x}e^{-si\theta_{\vec{k}}}&=\left(\frac{k_x}{k}+si\frac{k_y}{k}\right)\left(\frac{k-k_x\frac{k_x}{k}}{k^2}-i\frac{-sk_y\frac{k_x}{k}}{k^2}\right)\notag\\
    &=+si\frac{k_y}{k^2}\left(\frac{k_x}{k}+si\frac{k_y}{k}\right)\left(\frac{k_x}{k}-si\frac{k_y}{k}\right)\notag\\
    &=+s\frac{i}{k}\sin\theta_{\vec{k}},
\end{align}
\begin{align}
    e^{+si\theta_{\vec{k}}}\partial_{k_y}e^{-si\theta_{\vec{k}}}&=\left(\frac{k_x}{k}+si\frac{k_y}{k}\right)\left(\frac{-k_x\frac{k_y}{k}}{k^2}-si\frac{k-k_y\frac{k_y}{k}}{k^2}\right)\notag\\
    &=-si\frac{k_x}{k^2}\left(\frac{k_x}{k}+si\frac{k_y}{k}\right)\left(\frac{k_x}{k}-si\frac{k_y}{k}\right)\notag\\
    &=-s\frac{i}{k}\cos\theta_{\vec{k}},
\end{align} 
we attain 
\begin{align} \label{vdfvk1}
    \vec{\mathcal{R}}^{++}_{\vec{k}s}&=i\cos^2\frac{\Theta_{k}}{2}e^{+si\theta_{\vec{k}}}\partial_{\vec{k}}e^{-si\theta_{\vec{k}}}\\
    &=\frac{1}{2k}(1+\cos\Theta_{k})\left[-s\sin \theta_{\vec{k}},s\cos \theta_{\vec{k}}\right],\notag
\end{align}
\begin{align} 
    \vec{\mathcal{R}}^{+-}_{\vec{k}s}
    &=\frac{i}{2}\sin\Theta_{k}e^{+si\theta_{\vec{k}}}\partial_{\vec{k}}e^{-si\theta_{\vec{k}}}+\frac{i}{2}\partial_{\vec{k}}\Theta_{k}\\
    &=\frac{1}{2k}\sin\Theta_{k}\left[-s\sin \theta_{\vec{k}},s\cos \theta_{\vec{k}}\right]+\frac{i}{2}\partial_{\vec{k}}\Theta_{k},\notag 
\end{align}
\begin{align} 
    \vec{\mathcal{R}}^{-+}_{\vec{k}s}
    &=\frac{i}{2}\sin\Theta_{k}e^{+si\theta_{\vec{k}}}\partial_{\vec{k}}e^{-si\theta_{\vec{k}}}-\frac{i}{2}\partial_{\vec{k}}\Theta_{k}\\
    &=\frac{1}{2k}\sin\Theta_{k}\left[-s\sin \theta_{\vec{k}},s\cos \theta_{\vec{k}}\right]-\frac{i}{2}\partial_{\vec{k}}\Theta_{k},\notag 
\end{align}
\begin{align} \label{vdfvk4}
    \vec{\mathcal{R}}^{--}_{\vec{k}s}&=i\sin^2\frac{\Theta_{k}}{2}e^{+si\theta_{\vec{k}}}\partial_{\vec{k}}e^{-si\theta_{\vec{k}}}\\
    &=\frac{1}{2k}(1-\cos\Theta_{k})\left[-s\sin \theta_{\vec{k}},s\cos \theta_{\vec{k}}\right].\notag 
\end{align}
The above Berry connection (\ref{vdfvk1}-\ref{vdfvk4}) can be written into the matrix form in eigen basis 
\begin{align} \label{vfmvkdvfx}
    \mathcal{R}_{x,\vec{k}s}&=-\frac{s\sin \theta_{\vec{k}}}{2k}\left(\eta^o_{\vec{k}}+\cos\Theta_{k}\eta^z_{\vec{k}}+\sin\Theta_{k}\eta^x_{\vec{k}}\right)\\
    &-\frac{A(M+Bk^2)}{2\mathcal{E}^2_k}\cos\theta_{\vec{k}}\eta^y_{\vec{k}},\notag
\end{align}
\begin{align} \label{vfmvkdvfy}
    \mathcal{R}_{y,\vec{k}s}&=+\frac{s\cos \theta_{\vec{k}}}{2k}\left(\eta^o_{\vec{k}}+\cos\Theta_{k}\eta^z_{\vec{k}}+\sin\Theta_{k}\eta^x_{\vec{k}}\right)\\
    &-\frac{A(M+Bk^2)}{2\mathcal{E}^2_k}\sin\theta_{\vec{k}}\eta^y_{\vec{k}}.\notag 
\end{align}
Therefore, the off-diagonal components of the Berry connection are given by 
\begin{align} \label{vfgbgbx}
    \mathcal{R}^{\bar{\eta}\eta}_{x,\vec{k}s}=-s\frac{A}{2\mathcal{E}_{k}}\sin \theta_{\vec{k}}-i\eta\frac{A(M+Bk^2)}{2\mathcal{E}^2_k}\cos\theta_{\vec{k}},
\end{align}
\begin{align} \label{vfgbgby}
    \mathcal{R}^{\bar{\eta}\eta}_{y,\vec{k}s}=+s\frac{A}{2\mathcal{E}_{k}}\cos \theta_{\vec{k}}-i\eta\frac{A(M+Bk^2)}{2\mathcal{E}^2_k}\sin\theta_{\vec{k}}.
\end{align}

\subsection{Derivations of the velocity operators}

The velocity operator is defined by
\begin{align} \label{vfdvfjd}
    \hat{v}_{is}=\frac{\partial }{\hbar\partial k_i} \hat{\mathcal{H}}^e_0(\vec{r}_s),
\end{align}
i.e., 
\begin{align} \label{vfdvfjdx}
     \hat{v}_{xs}=s\frac{A}{\hbar}\hat{\sigma}_x-\frac{2Bk}{\hbar}\cos\theta_{\vec{k}}\hat{\sigma}_z,
\end{align}
\begin{align} \label{vfdvfjdy}
    \hat{v}_{ys}=\frac{A}{\hbar}\hat{\sigma}_y-\frac{2Bk}{\hbar}\sin\theta_{\vec{k}}\hat{\sigma}_z, 
\end{align}
which are diagonal in momentum space. Then, we express the velocity operators in terms of  eigenstate basis [Eqs.~\eqref{fdvefk1} and \eqref{fdvefk2}]. Noting that
\begin{align} 
    \sigma^x_{\vec{k}s}=-s\cos\Theta_{k}\cos\theta_{\vec{k}}\eta^x_{\vec{k}}+\sin\theta_{\vec{k}}\eta^y_{\vec{k}}+s\sin\Theta_{k}\cos\theta_{\vec{k}}\eta^z_{\vec{k}},
\end{align}
\begin{align} 
\sigma^y_{\vec{k}s}=-\cos\Theta_{k}\sin\theta_{\vec{k}}\eta^x_{\vec{k}}-s\cos\theta_{\vec{k}}\eta^y_{\vec{k}}+\sin\Theta_{k}\sin\theta_{\vec{k}}\eta^z_{\vec{k}},
\end{align}
\begin{align} 
    \sigma^z_{\vec{k}s}=\sin\Theta_{k}\eta^x_{\vec{k}}+\cos\Theta_{k}\eta^z_{\vec{k}},
\end{align} 
the velocity operators \eqref{vfdvfjdx} and \eqref{vfdvfjdy}, in eigen basis, are given by 
\begin{align} \label{fyqvx}
    \hat{v}^x_{\vec{k}s} 
    &= \left(-\frac{A}{\hbar}\cos\Theta_{k}\cos\theta_{\vec{k}}-\frac{2Bk}{\hbar}\sin\Theta_{k}\cos\theta_{\vec{k}}\right)\eta^x_{\vec{k}}\notag\\
    &+ s\frac{A}{\hbar}\sin\theta_{\vec{k}}\eta^y_{\vec{k}}\\
    &+\left(\frac{A}{\hbar}\sin\Theta_{k}\cos\theta_{\vec{k}}-\frac{2Bk}{\hbar}\cos\Theta_{k}\cos\theta_{\vec{k}}\right)\eta^z_{\vec{k}},\notag
\end{align}
\begin{align} \label{fyqvy}
    \hat{v}^y_{\vec{k}s}
    &=\left(-\frac{A}{\hbar}\cos\Theta_{k}\sin\theta_{\vec{k}}-\frac{2Bk}{\hbar}\sin\Theta_{k}\sin\theta_{\vec{k}}\right)\eta^x_{\vec{k}}\notag \\
    &- \frac{A}{\hbar}s\cos\theta_{\vec{k}}\eta^y_{\vec{k}}\\
    &+ \left(\frac{A}{\hbar}\sin\Theta_{k}\sin\theta_{\vec{k}}-\frac{2Bk}{\hbar}\cos\Theta_{k}\sin\theta_{\vec{k}}\right)\eta^z_{\vec{k}}.\notag
\end{align}
Comparing Eqs.~\eqref{vfgbgbx}, \eqref{vfgbgby} with Eqs. \eqref{fyqvx}and \eqref{fyqvy}, 
we get the following relationship 
\begin{align} \label{gbadjfjfkfg}
    \mathcal{R}^{\bar{\eta}\eta}_{j,\vec{k}s}=\frac{\hbar}{i(\epsilon_{\vec{k}\bar{\eta}}-\epsilon_{\vec{k}\eta})}\left\{\hat{v}^{0}_{j,\vec{k}s}\right\}^{\bar{\eta}\eta}.
\end{align}

\section{Derivations of charge and spin conductivity} \label{derivationcurrent}

The charge current \textit{density} for each spin block, when itinerant electrons exist in electric field-induced superposition states, can be defined from density matrix $\varrho_{\vec{k}}$ as follows
\begin{align} \label{fvdfkafvf}
    J^{s}_{i}=\frac{e}{V}\sum_{\vec{\vec{k}}\eta}\hat{v}^{\eta\eta}_{i,\vec{k}s} \varrho^{\eta\eta}_{\vec{k}s}+\frac{e}{V}\sum_{\vec{\vec{k}}\eta}\hat{v}_{i,\vec{k}s}^{\eta\bar{\eta}} \varrho^{\bar{\eta}\eta}_{\vec{k}s},
\end{align}
with $i=(x,y)$  and $s=\uparrow/\downarrow$, where $V$ is area of the sample. The first term describes the contribution from that diagonal density matrix ($\varrho^{\eta\eta}_{\vec{k}s}$) that quantifies the quantum distribution of itinerant electrons, while the second term contains the contribution from off-diagonal component ($\varrho^{\bar{\eta}\eta}_{\vec{k}s}$) of the density matrix that quantifies the quantum coherence of itinerant electrons~\footnote{For a superposition state $\psi_{\vec{k}}=a_{\vec{k}}\vert \vec{k}+\rangle+b_{\vec{k}}\vert \vec{k}+\rangle$, the corresponding density matrix is $\varrho_{\vec{k}}\equiv \psi_{\vec{k}}\psi^{\dagger}_{\vec{k}}=[\vert a_{\vec{k}} \vert^2 ,a_{\vec{k}}b^{*}_{\vec{k}} ; a^{*}_{\vec{k}}b_{\vec{k}}, \vert b_{\vec{k}} \vert^2]$. Obviously, the diagonal components  of density matrix describe the distribution of itinerant electron, while the off-diagonal component quantify the quantum coherence and acquire finite value only when electrons occupy multiple quantum states simultaneously.}. In the absence of external electric field, we attain equilibrium velocity operator $\hat{v}^{i}_{\vec{k}}$ and equilibrium distribution function $\varrho^{\eta\eta'}_{\vec{k}s,0}=\delta_{\eta\eta'}f_{k\eta}$. Then, the above current density \eqref{fvdfkafvf} vanishes, i.e., $ J^{s}_{i}=0$.

\subsection{Conductivities from Berry curvature}

\subsubsection{Derivations of the Berry curvatures}

The the Berry curvature can be expressed as following:
\begin{align} \label{gbsgbgf}
   \Omega^z_{\vec{k}s\eta} = -\eta\frac{1}{2} \frac{\mathbf{d} \cdot (\partial_{k_x} \mathbf{d} \times \partial_{k_y} \mathbf{d})}{|\mathbf{d}|^3}.
\end{align}
Based on our Bernevig–Hughes–Zhang Hamiltonian, i.e.,  
\begin{align} \label{fvagbg}
    H_s(\mathbf{k}) = sAk^x\hat{\sigma}_x+Ak_y\hat{\sigma}^y+(M-Bk^2)\hat{\sigma}^z,
\end{align}
we can have
\begin{align}
    \mathbf{d}(\mathbf{k})=(sAk^x,Ak_y,M-Bk^2).
\end{align}
from that, we compute
\begin{align}
    \frac{\partial \mathbf{d}}{\partial k_x} = (sA, 0, -2Bk_x), 
\end{align}
\begin{align}
\frac{\partial \mathbf{d}}{\partial k_y} = (0, A, -2Bk_y).
\end{align}
Then:
\begin{align}
    \frac{\partial \mathbf{d}}{\partial k_x} \times \frac{\partial \mathbf{d}}{\partial k_y} =
\begin{vmatrix}
\hat{x} & \hat{y} & \hat{z} \\
sA & 0 & -2Bk_x \\
0 & A & -2Bk_y
\end{vmatrix}
= \left( 2ABk_x, 2sABk_y, sA^2 \right).
\end{align}
Therefore, the Berry curvature \eqref{gbsgbgf} becomes
\begin{align}
    \Omega^z_{\vec{k}s\eta} 
    &= -s\eta\frac{A^2}{2} \frac{M+Bk^2}{\left[ A^2k^2+(M-Bk^2)^2 \right]^{3/2}}.
\end{align}

\subsubsection{Conductivity from Berry curvature} \label{Berrycurvaturederivation}

Let us begin with the intrinsic mechanism from the Berry curvature. In modern semiclassical theory, drift velocity appears in the equations of motion, but it's corrected by quantum geometric effects such as Berry curvature $\vec{\Omega}_{\vec{k}s\eta}=\vec{\nabla}_{\vec{k}}\times \vec{\mathcal{R}}^{\eta\eta}_{\vec{k}s} $ by the  anomalous velocity We have anomalous velocity  in transverse direction 
\begin{align} \label{dfvdlfl}
    \hat{v}^{\text{av},\eta\eta'}_{y,\vec{k}s}=\frac{e}{\hbar}\delta_{\eta\eta'}\Omega^{z}_{\vec{k}s\eta}E_{x},
\end{align}
where we have used the expressions of the Berry curvature as follows 
\begin{align}
    \Omega^z_{\vec{k}s\eta} 
= -s\eta\frac{A^2}{2} \frac{M+Bk^2}{\left[ A^2k^2+(M-Bk^2)^2 \right]^{3/2}}.
\end{align}
The  charge Hall current for each spin block, defined from the anomalous velocity \eqref{dfvdlfl}, are  given by
\begin{align}
    J^{s}_{y}=\frac{e}{2V}\sum_{\vec{\vec{k}}\eta}\left\{\hat{v}^{\text{av}}_{y,\vec{k}s}\right\}^{\eta\eta} f^{}_{\vec{k}\eta}.
\end{align}
By substitution of anomalous velocity, we reach 
\begin{align}
    J^{s}_{y}=-s\frac{e^2}{\hbar V}\sum_{\vec{\vec{k}}\eta} \eta\frac{A^2}{2} \frac{M+Bk^2}{\left[ A^2k^2+(M-Bk^2)^2 \right]^{3/2}} f^{}_{\vec{k}\eta}E_{x}.
\end{align}
The associated conductivity, at zero temperature, is given by 
\begin{align}
    \sigma^s_H=s\frac{e^2}{\hbar V}\sum_{\vec{\vec{k}}}\frac{A^2}{2} \frac{M+Bk^2}{\left[ A^2k^2+(M-Bk^2)^2 \right]^{3/2}} .
\end{align}
Then, we replace the sum over $\vec{k}$ with integral, i.e., 
\begin{align} \label{fvfvkgdmk}
    \frac{1}{V}\sum_{\vec{k}}\rightarrow \frac{1}{(2\pi)^2}\int^{\infty}_0 k dk \int^{2\pi}_0 d\theta_{\vec{k}},
\end{align}
leading to
\begin{align}\label{fyqsigmaH}
    \sigma^s_H=s\frac{e^2}{2h }\int^{\infty}_0 k dk A^2 \frac{M+Bk^2}{\left[ A^2k^2+(M-Bk^2)^2 \right]^{3/2}} .
\end{align}
Next, we derive the above integrals. We attain 
\begin{align}
    MA^2\int_0^{\infty}dk\frac{k}{\mathcal{E}^3_k}&=\frac{M}{2}\int_0^{\infty}dk^2\frac{1}{\{k^2+[M-(B/A^2)k^2]^2\}^{3/2}}\notag \\
    &=\frac{M}{2}\int_0^{\infty}dx\frac{1}{\{x+[M-(B/A^2)x]^2\}^{3/2}}, 
\end{align}
\begin{align}
    BA^2\int^{\infty}_0 dk\frac{k^3}{\mathcal{E}^3_{k}}&=\frac{B}{2A^2}\int_0^{\infty}dk^2\frac{k^2}{\{k^2+[M-(B/A^2)k^2]^2\}^{3/2}}\notag \\
    &=\frac{B}{2A^2}\int_0^{\infty}dx\frac{x}{\{x+[M-(B/A^2)x]^2\}^{3/2}}.
\end{align}
Using the following integral rules
\begin{align}
   \int_0^{\infty}dx\frac{1}{\{ax^2+bx+c\}^{3/2}}=\frac{2(2ax+b)}{(4ac-b^2)\sqrt{ax^2+bx+c}}
   \end{align}
\begin{align}
   \int_0^{\infty}dx\frac{x}{\{ax^2+bx+c\}^{3/2}}=\frac{-2(2c+bx)}{(4ac-b^2)\sqrt{ax^2+bx+c}}
   \end{align}  
we reach
\begin{align} \label{fvfvdk1}
    \frac{M}{2}&\int_0^{\infty}dx\frac{1}{\{x+[M-(B/A^2)x]^2\}^{3/2}}\\
    &=\frac{M}{\vert M\vert+2M^2\vert B\vert/A^2[1-\text{sign}(MB)] } \notag\\
    &=\left\{\begin{matrix*}
       \text{sign}(M), & \text{sign}(MB)>0\\
       \frac{\text{sign}(M)}{1+4\vert M\vert\vert B\vert/A^2 }, & \text{sign}(MB)<0
    \end{matrix*}\right.,\notag
\end{align}
\begin{align} \label{fvfvdk3}
    \frac{B}{2A^2}&\int_0^{\infty}dx\frac{x}{\{x+[M-(B/A^2)x]^2\}^{3/2}}\\
    &=\frac{B/A^2}{\vert B\vert/A^2+2\vert M\vert B^2/A^4[1-\text{sign}(MB)] }\notag \\
    &=\left\{\begin{matrix*}
       \text{sign}(B), & \text{sign}(MB)>0\\
       \frac{\text{sign}(B)}{1+4\vert M\vert\vert B\vert/A^2 }, & \text{sign}(MB)<0
    \end{matrix*}\right.,\notag 
\end{align}
By substitution of Eqs.~\eqref{fvfvdk1} and \eqref{fvfvdk3}, the Hall conductivity \eqref{fyqsigmaH} becomes 
\begin{align} 
     \sigma^{s}_{H}=s\frac{e^2}{h}\left\{\begin{matrix*}
        \text{sign}(M), & \text{sign}(MB)>0\\
       0, & \text{sign}(MB)<0
    \end{matrix*}\right..
\end{align}
Then, we attain the spin Hall conductivity for the quantum spin Hall effect  
\begin{align} \label{fdfvdfkk}
   \sigma^{z}_{H}=\sigma^{\uparrow}_H-\sigma^{\downarrow}_H=\frac{2e^2}{h}\left\{\begin{matrix*}
       \text{sign}(M), & \text{sign}(MB)>0\\
       0, & \text{sign}(MB)<0
    \end{matrix*}\right..
\end{align}
While the charge Hall conductivity vanishes as follows
\begin{align} \label{gflblgl}
    \sigma^{o}_{H}=\sigma^{\uparrow}_H+\sigma^{\downarrow}_H=0,
\end{align}
because the anomalous velocity from the Berry curvature is always perpendicular to $\vec{E}$.

\subsection{Charge conductivity from quantum coherence}
The charge current density that arises from the off-diagonal density matrix, is expressed as 
\begin{align} \label{jxoofffd}
    J^{s}_i=\frac{e}{V}\sum_{\vec{\vec{k}}s\eta}(\hat{v}^{i}_{\vec{k}s})^{\eta\bar{\eta}} \delta\varrho^{\bar{\eta}\eta}_{\vec{k}s}.
\end{align}
The  off-diagonal components of the longitudinal and transverse charge  current operators are, respectively,  given by according to Eqs.~\eqref{fyqvx} and \eqref{fyqvy}
\begin{align} 
   (\hat{v}^{x}_{\vec{k}s})^{\eta\bar{\eta}}&=-\frac{A(M+Bk^2)}{\hbar \mathcal{E}_{k}}\cos\theta_{\vec{k}}-is\eta\frac{A}{\hbar}\sin\theta_{\vec{k}},
\end{align}
\begin{align}
    (\hat{v}^{y}_{\vec{k}s})^{\eta\bar{\eta}}&= -\frac{A(M+Bk^2)}{\hbar \mathcal{E}_{k}}\sin\theta_{\vec{k}}+is\eta\frac{A}{\hbar}\cos\theta_{\vec{k}}. 
\end{align}
The Berry connection is given by 
\begin{align} 
    \mathcal{R}^{\bar{\eta}\eta}_{x,\vec{k}s}=-s\frac{A}{2\mathcal{E}_{k}}\sin \theta_{\vec{k}}-i\eta\frac{A(M+Bk^2)}{2\mathcal{E}^2_k}\cos\theta_{\vec{k}},
\end{align}
\begin{align} 
    \mathcal{R}^{\bar{\eta}\eta}_{y,\vec{k}s}=+s\frac{A}{2\mathcal{E}_{k}}\cos \theta_{\vec{k}}-i\eta\frac{A(M+Bk^2)}{2\mathcal{E}^2_k}\sin\theta_{\vec{k}}.
\end{align}
We separate the contributions from ordinary and anomalous density matrix 
\begin{align} 
    J^{s}_{x}=\sigma^{s}_{L,\Vert}E^x+\sigma^{s}_{L,\perp}E^x,
\end{align}
\begin{align} 
    J^{s}_{y}=\sigma^{s}_{H,\Vert}E^x+\sigma^{s}_{H,\perp}E^x,
\end{align}
with
\begin{align} \label{jxaoffx}
    \sigma^{s}_{L,\Vert/\perp}=\frac{e}{V}\sum_{\vec{\vec{k}}\eta}(\hat{v}_{\vec{k}s}^{x})^{\eta\bar{\eta}} \delta\varrho^{\bar{\eta}\eta}_{\vec{k}s,\Vert/\perp}/E^x,
\end{align}
\begin{align} \label{jxaoffy}
    \sigma^{s}_{H,\Vert/\perp}=\frac{e}{V}\sum_{\vec{\vec{k}}\eta}(\hat{v}_{\vec{k}s}^{y})^{\eta\bar{\eta}} \delta\varrho^{\bar{\eta}\eta}_{\vec{k}s,\Vert/\perp}/E^x.
\end{align}

\subsubsection{Case i: in the absence of quantum decoherence}

In this case, we have off-diagonal density matrix
\begin{align} \label{fvgbhbgt}
    \delta\varrho^{\bar{\eta}\eta}_{\vec{k}s,\Vert}= -e\frac{ (f_{k+}- f_{k-})}{2\mathcal{E}_{k}} \mathcal{R}^{x,\bar{\eta}\eta}_{\vec{k}s}E^x,
\end{align}
\begin{align}
    \delta\varrho^{\bar{\eta}\eta}_{\vec{k}s,\perp}=0.
\end{align}
By means of Eq. \eqref{fvgbhbgt}, Eqs. \eqref{jxaoffx} and \eqref{jxaoffy} become
\begin{widetext}
\begin{align} \label{jxaroffx}
    \sigma^{s}_{L,\Vert}=-\frac{e^2}{V}\sum_{\vec{\vec{k}}\eta}\frac{f_{k+}- f_{k-}}{2\mathcal{E}_{k}}\left[-\frac{A(M+Bk^2)}{\hbar \mathcal{E}_{k}}\cos\theta_{\vec{k}}-is\eta\frac{A}{\hbar}\sin\theta_{\vec{k}}\right]  \left[-s\frac{A}{2\mathcal{E}_{k}}\sin \theta_{\vec{k}}-i\eta\frac{A(M+Bk^2)}{2\mathcal{E}^2_k}\cos\theta_{\vec{k}}\right],
\end{align}
\begin{align} \label{jxaroffy}
    \sigma^{s}_{H,\Vert}=-\frac{e^2}{V}\sum_{\vec{\vec{k}}\eta}\frac{ f_{k+}- f_{k-}}{2\mathcal{E}_{k}}\left[-\frac{A(M+Bk^2)}{\hbar \mathcal{E}_{k}}\sin\theta_{\vec{k}}+is\eta\frac{A}{\hbar}\cos\theta_{\vec{k}}\right]  \left[-s\frac{A}{2\mathcal{E}_{k}}\sin \theta_{\vec{k}}-i\eta\frac{A(M+Bk^2)}{2\mathcal{E}^2_k}\cos\theta_{\vec{k}}\right],
\end{align}
\end{widetext}
i.e.,
\begin{align} \label{jxaroffx}
    \sigma^{s}_{L,\Vert}=0,
\end{align}
\begin{align} \label{jxaroffy}
    \sigma^{s}_{H,\Vert}&=-s\frac{e^2}{V}\sum_{\vec{\vec{k}}\eta}\frac{ (f_{k+}- f_{k-})}{2\mathcal{E}_{k}}\\
    &\times\left[\left(M+Bk^2\right)\frac{A^2}{2\hbar\mathcal{E}^2_{k}}\sin^2\theta_{\vec{k}}+\frac{A^2(M+Bk^2)}{2\hbar\mathcal{E}^2_k}\cos^2\theta_{\vec{k}}\right]. \notag 
\end{align}
Then, we replace the sum over $\vec{k}$ with integral as follows 
\begin{align} \label{fvfvkgmk}
    \frac{1}{V}\sum_{\vec{k}}\rightarrow \frac{1}{(2\pi)^2}\int^{\infty}_0 k dk \int^{2\pi}_0 d\theta_{\vec{k}},
\end{align}
leading to
\begin{align} \label{rjxaroffy}
    \sigma^{s}_{H,\Vert}=-s\frac{e^2}{2h}\int^{\infty}_0 kdk\frac{ (f_{k+}- f_{k-})}{\mathcal{E}_{k}}\frac{A^2(M+Bk^2)}{\mathcal{E}^2_{k}}.
\end{align}
For energy level inside the gap, we attain 
\begin{align} \label{rjxaroffyrav}
    \sigma^{s}_{H,\Vert}=+s\frac{e^2}{2h}\int^{\infty}_0 dk\frac{A^2(M+Bk^2)k}{\mathcal{E}^3_{k}}.
\end{align}
This result is same to Eq.~\eqref{fyqsigmaH} from the Berry curvature.
Hence, we reach
\begin{align} 
     \sigma^{s}_{H,\Vert}=s\frac{e^2}{h}\left\{\begin{matrix*}
       +\text{sign}(M), & \text{sign}(MB)>0\\
       0, & \text{sign}(MB)<0
    \end{matrix*}\right..
\end{align}
Then, we attain the longitudinal charge (transverse spin) conductivity  for the quantum magnetoresistance (quantum spin Hall) effect  
\begin{align} \label{fdfvdfkko}
   \sigma^{o}_{L,\Vert}=0
\end{align}
\begin{align} \label{fdfvdfkkz}
   \sigma^{z}_{H,\Vert}=\sigma^{\uparrow}_{H,\Vert}-\sigma^{\downarrow}_{H,\Vert}=+\frac{2e^2}{h}\left\{\begin{matrix*}
       \text{sign}(M), & \text{sign}(MB)>0\\
       0, & \text{sign}(MB)<0
    \end{matrix*}\right..
\end{align}
While the charge Hall conductivity vanishes as follows 
\begin{align} \label{gvgakbfk}
    \sigma^{o}_{H,\Vert}=\sigma^{\uparrow}_{H,\Vert}+\sigma^{\downarrow}_{H,\Vert}=0.
\end{align}
These results, i.e., Eqs. \eqref{fdfvdfkkz} and \eqref{gvgakbfk}, are consistent with the ones of the Berry curvature, i.e.,  Eqs. \eqref{fdfvdfkk} and \eqref{gflblgl}. 

\subsubsection{Case ii: in the presence of quantum decoherence}
In this case, we have off-diagonal density matrix
\begin{align} \label{ybgbv}
\delta\varrho^{\bar{\eta}\eta}_{\vec{k}s,\Vert}=-\frac{e}{\hbar}\tau^{\bar{\eta}\eta}_{ks,\Vert}(f_{k\bar{\eta}}- f_{k\eta})\mathcal{R}^{x,\bar{\eta}\eta}_{\vec{k}s}E^x,
\end{align}
\begin{align} \label{ybgbp}
\delta\varrho^{\bar{\eta}\eta}_{\vec{k}s,\perp}=+\frac{e}{\hbar}\tau^{\bar{\eta}\eta}_{ks,\perp}(f_{k\bar{\eta}}- f_{k\eta})\mathcal{R}^{y,\bar{\eta}\eta}_{\vec{k}s}E^x,
\end{align}
with
\begin{align}
     \frac{\tau^{\bar{\eta}\eta}_{ks,\Vert}}{\hbar}=\frac{(\epsilon_{k\bar{\eta}}-\epsilon_{k\eta})-i\Gamma_{k}}{[(\epsilon_{k\bar{\eta}}-\epsilon_{k\eta})-i\Gamma_{k}]^2+(\Gamma^a_{k})^2},
\end{align}
\begin{align}
     \frac{\tau^{\bar{\eta}\eta}_{ks,\perp}}{\hbar}=\frac{s\eta\Gamma^a_{k}}{[(\epsilon_{k\bar{\eta}}-\epsilon_{k\eta})-i\Gamma_{k}]^2+(\Gamma^a_{k})^2},
\end{align}
which satisfy the following conjugationg relations 
\begin{align} \label{vdvgkv1}
   (f_{k-}- f_{k+}) \tau^{-+}_{ks,\Vert}
   &=\left[(f_{k+}- f_{k-})\tau^{+-}_{ks,\Vert}\right]^*,
\end{align}
\begin{align} \label{vdvgkv2}
   (f_{k-}- f_{k+})\tau^{-+}_{ks,\perp}
   &=\left[(f_{k+}- f_{k-})\tau^{+-}_{ks,\perp}\right]^*.
\end{align}
By means of Eqs. \eqref{ybgbv} and \eqref{ybgbp}, Eqs. \eqref{jxaoffx} and \eqref{jxaoffy} become
\begin{widetext}
\begin{align} \label{sxJoxiot}
    \sigma^{s}_{L,\Vert}=-\frac{e^2}{V}\sum_{\vec{\vec{k}}\eta}(f_{k\bar{\eta}}- f_{k\eta})\frac{\tau^{\bar{\eta}\eta}_{ks,\Vert}}{\hbar}\left[-\frac{A(M+Bk^2)}{\hbar \mathcal{E}_{k}}\cos\theta_{\vec{k}}-is\eta\frac{A}{\hbar}\sin\theta_{\vec{k}}\right]  \left[-s\frac{A}{2\mathcal{E}_{k}}\sin \theta_{\vec{k}}-i\eta\frac{A(M+Bk^2)}{2\mathcal{E}^2_k}\cos\theta_{\vec{k}}\right],
\end{align}
\begin{align} \label{sxJoyiot}
    \sigma^{s}_{L,\perp}=+\frac{e^2}{V}\sum_{\vec{\vec{k}}\eta}(f_{k\bar{\eta}}- f_{k\eta})\frac{\tau^{\bar{\eta}\eta}_{ks,\perp}}{\hbar}\left[-\frac{A(M+Bk^2)}{\hbar \mathcal{E}_{k}}\cos\theta_{\vec{k}}-is\eta\frac{A}{\hbar}\sin\theta_{\vec{k}}\right]  \left[+s\frac{A}{2\mathcal{E}_{k}}\cos \theta_{\vec{k}}-i\eta\frac{A(M+Bk^2)}{2\mathcal{E}^2_k}\sin\theta_{\vec{k}}\right],
\end{align}
\begin{align} \label{syJoxiot}
    \sigma^{s}_{H,\Vert}=-\frac{e^2}{V}\sum_{\vec{\vec{k}}\eta}(f_{k\bar{\eta}}- f_{k\eta})\frac{\tau^{\bar{\eta}\eta}_{ks,\Vert}}{\hbar} \left[-\frac{A(M+Bk^2)}{\hbar \mathcal{E}_{k}}\sin\theta_{\vec{k}}+is\eta\frac{A}{\hbar}\cos\theta_{\vec{k}}\right] \left[-s\frac{A}{2\mathcal{E}_{k}}\sin \theta_{\vec{k}}-i\eta\frac{A(M+Bk^2)}{2\mathcal{E}^2_k}\cos\theta_{\vec{k}}\right],
\end{align}
\begin{align} \label{syJoyiot}
    \sigma^{s}_{H,\perp}=+\frac{e^2}{V}\sum_{\vec{\vec{k}}\eta}(f_{k\bar{\eta}}- f_{k\eta})\frac{\tau^{\bar{\eta}\eta}_{ks,\perp}}{\hbar}\left[-\frac{A(M+Bk^2)}{\hbar \mathcal{E}_{k}}\sin\theta_{\vec{k}}+is\eta\frac{A}{\hbar}\cos\theta_{\vec{k}}\right]  \left[+s\frac{A}{2\mathcal{E}_{k}}\cos \theta_{\vec{k}}-i\eta\frac{A(M+Bk^2)}{2\mathcal{E}^2_k}\sin\theta_{\vec{k}}\right],
\end{align}
\end{widetext}
i.e.,
\begin{align} \label{jxaroffxt}
    \sigma^{s}_{L,\Vert}&=-\frac{e^2}{V}\sum_{\vec{\vec{k}}\eta}(f_{k\bar{\eta}}- f_{k\eta})\frac{\tau^{\bar{\eta}\eta}_{ks,\Vert}}{\hbar}\\
    &\times \left[i\eta\frac{A^2(M+Bk^2)^2}{2\hbar \mathcal{E}^3_k}\cos^2\theta_{\vec{k}}+i\eta\frac{A^2}{2\hbar\mathcal{E}_k}\sin^2\theta_{\vec{k}}\right],\notag
\end{align}
\begin{align} \label{jxaroffxtp}
    \sigma^{s}_{L,\perp}=-s\frac{e^2}{V}\sum_{\vec{\vec{k}}\eta}(f_{k\bar{\eta}}- f_{k\eta})\frac{\tau^{\bar{\eta}\eta}_{ks,\perp}}{\hbar}\left[\frac{A^2(M+Bk^2)}{2\hbar\mathcal{E}^2_k}\right],
\end{align}
\begin{align} \label{jxaroffyt}
    \sigma^{s}_{H,\Vert}&=-s\frac{e^2}{V}\sum_{\vec{\vec{k}}\eta}(f_{k\bar{\eta}}- f_{k\eta})\frac{\tau^{\bar{\eta}\eta}_{ks,\Vert}}{\hbar} \left[\frac{A^2(M+Bk^2)}{2\hbar\mathcal{E}^2_k}\right],
\end{align}
\begin{align} \label{jxaroffytp}
    \sigma^{s}_{H,\perp}&=\frac{e^2}{V}\sum_{\vec{\vec{k}}\eta}(f_{k\bar{\eta}}- f_{k\eta})\frac{\tau^{\bar{\eta}\eta}_{ks,\perp}}{\hbar}\\
    &\times\left[i\eta\frac{A^2(M+Bk^2)^2}{2\hbar\mathcal{E}^3_k}\sin^2\theta_{\vec{k}}+i\frac{A^2}{2\hbar\mathcal{E}_{k}}\cos^2\theta_{\vec{k}}\right].\notag
\end{align}
Integrating over $\theta_{\vec{k}}$
leads to 
\begin{align} \label{4jxaroffxt}
    \sigma^{s}_{L,\Vert}&=-\frac{e^2}{2h}\sum_{\eta}\int^{\infty}_{0}kdk(f_{k\bar{\eta}}- f_{k\eta})\frac{\tau^{\bar{\eta}\eta}_{ks,\Vert}}{\hbar}\\
    &\times \left[i\eta\frac{A^2}{2 \mathcal{E}_k}+i\eta \frac{A^2(M+Bk^2)^2}{2\mathcal{E}^3_k}\right],\notag
\end{align}
\begin{align} \label{4jxaroffxtp}
    \sigma^{s}_{L,\perp}&=-s\frac{e^2}{2h}\sum_{\eta}\int^{\infty}_{0}kdk(f_{k\bar{\eta}}- f_{k\eta})\frac{\tau^{\bar{\eta}\eta}_{ks,\perp}}{\hbar}\\
    &\times\left[\frac{A^2(M+Bk^2)}{\mathcal{E}^2_k}\right],\notag
\end{align}
\begin{align} \label{4jxaroffyt}
    \sigma^{s}_{H,\Vert}&=-s\frac{e^2}{2h}\sum_{\eta}\int^{\infty}_{0}kdk(f_{k\bar{\eta}}- f_{k\eta})\frac{\tau^{\bar{\eta}\eta}_{ks,\Vert}}{\hbar}\\
    &\times\left[\frac{A^2(M+Bk^2)}{\mathcal{E}^2_k}\right],\notag
\end{align}
\begin{align} \label{4jxaroffytp}
    \sigma^{s}_{H,\perp}&=\frac{e^2}{2h}\sum_{\eta}\int^{\infty}_{0}kdk(f_{k\bar{\eta}}- f_{k\eta})\frac{\tau^{\bar{\eta}\eta}_{ks,\perp}}{\hbar}\\
    &\times\left[i\eta\frac{A^2(M+Bk^2)^2}{2\mathcal{E}^3_k}+i\eta\frac{A^2}{2\mathcal{E}_{k}}\right],\notag
\end{align}
Finally, we attain the expressions of the charge conductivities of each spin block in both longitudinal and transverse directions as follows
\begin{align} \label{f4jxaroffxttT}
    \sigma^{s}_{L,\Vert}(T)&=    -\frac{e^2}{2h}\int^{\infty}_{0}kdk \text{Im}\left\{\frac{\tau^{+-}_{ks,\Vert}}{\hbar}\right\}\left[\frac{A^2}{\mathcal{E}_k}+ \frac{A^2(M+Bk^2)^2}{\mathcal{E}^3_k}\right]\notag \\
    &\times \left[\frac{1}{(e^{\beta(\epsilon_{k+}-\epsilon_{F})} + 1)}-\frac{1}{(e^{\beta(\epsilon_{k-}-\epsilon_{F})} + 1)}\right], 
\end{align}
\begin{align} \label{f4jxaroffxtptT}
    \sigma^{s}_{L,\perp}(T)&=-s\frac{e^2}{h}\int^{\infty}_{0}kdk\text{Re}\left\{\frac{\tau^{+-}_{ks,\perp}}{\hbar}\right\}\left[\frac{A^2
    (M+Bk^2)}{\mathcal{E}^2_k}\right]\notag\\
    &\times \left[\frac{1}{(e^{\beta(\epsilon_{k+}-\epsilon_{F})} + 1)}-\frac{1}{(e^{\beta(\epsilon_{k-}-\epsilon_{F})} + 1)}\right],
\end{align}
\begin{align} \label{f4jxaroffyttT}
    \sigma^{s}_{H,\Vert}(T)&=-s\frac{e^2}{h}\int^{\infty}_{0}kdk\text{Re}\left\{\frac{\tau^{+-}_{ks,\Vert}}{\hbar}\right\}\left[\frac{A^2(M+Bk^2)}{\mathcal{E}^2_k}\right]\notag\\
    &\times \left[\frac{1}{(e^{\beta(\epsilon_{k+}-\epsilon_{F})} + 1)}-\frac{1}{(e^{\beta(\epsilon_{k-}-\epsilon_{F})} + 1)}\right],
\end{align}
\begin{align} \label{f4jxaroffytptT}
    \sigma^{s}_{H,\perp}(T)&=\frac{e^2}{2h}\int^{\infty}_{0}kdk \text{Im}\left\{\frac{\tau^{+-}_{ks,\perp}}{\hbar}\right\}\left[\frac{A^2}{\mathcal{E}_k}+\frac{A^2(M+Bk^2)^2}{\mathcal{E}^3_k} \right]\notag\\
    &\times \left[\frac{1}{(e^{\beta(\epsilon_{k+}-\epsilon_{F})} + 1)}-\frac{1}{(e^{\beta(\epsilon_{k-}-\epsilon_{F})} + 1)}\right].
\end{align}
At zero temperature, the above conductivities reduce to 
\begin{align} \label{f4jxaroffxtt}
    \sigma^{s}_{L,\Vert}=\frac{e^2}{2h}\int^{\infty}_{0}kdk \text{Im}\left\{\frac{\tau^{+-}_{ks,\Vert}}{\hbar}\right\}\left[\frac{A^2}{\mathcal{E}_k}+ \frac{A^2(M+Bk^2)^2}{\mathcal{E}^3_k}\right],
\end{align}
\begin{align} \label{f4jxaroffxtpt}
    \sigma^{s}_{L,\perp}=+s\frac{e^2}{h}\int^{\infty}_{0}kdk\text{Re}\left\{\frac{\tau^{+-}_{ks,\perp}}{\hbar}\right\}\left[\frac{A^2
    (M+Bk^2)}{\mathcal{E}^2_k}\right],
\end{align}
\begin{align} \label{f4jxaroffytt}
    \sigma^{s}_{H,\Vert}=+s\frac{e^2}{h}\int^{\infty}_{0}kdk\text{Re}\left\{\frac{\tau^{+-}_{ks,\Vert}}{\hbar}\right\}\left[\frac{A^2(M+Bk^2)}{\mathcal{E}^2_k}\right],
\end{align}
\begin{align} \label{f4jxaroffytpt}
    \sigma^{s}_{H,\perp}=-\frac{e^2}{2h}\int^{\infty}_{0}kdk \text{Im}\left\{\frac{\tau^{+-}_{ks,\perp}}{\hbar}\right\}\left[\frac{A^2}{\mathcal{E}_k}+\frac{A^2(M+Bk^2)^2}{\mathcal{E}^3_k} \right],
\end{align}
where we have used the conjugation relations \eqref{vdvgkv1} and \eqref{vdvgkv2}. Noting that $\epsilon^{+-}_{k,\Vert}$($\epsilon^{+-}_{ks,\perp}$) is independent of (linear in) $s$, we attain the following conductivities
\begin{align} \label{f4jxdaroffxttT}
    \sigma^{o}_{L,\Vert}(T)&=\sum_s\sigma^{s}_{L,\Vert}(T)\\
    &=-\frac{e^2}{h}\int^{\infty}_{0}kdk \text{Im}\left\{\frac{\tau^{+-}_{k,\Vert}}{\hbar}\right\}\left[\frac{A^2}{\mathcal{E}_k}+ \frac{A^2(M+Bk^2)^2}{\mathcal{E}^3_k}\right]\notag\\
    &\times \left[\frac{1}{(e^{\beta(\epsilon_{k+}-\epsilon_{F})} + 1)}-\frac{1}{(e^{\beta(\epsilon_{k-}-\epsilon_{F})} + 1)}\right],\notag 
\end{align}
\begin{align} \label{f4jdxdaroffxttT}
    \sigma^{z}_{L,\Vert}(T)=\sum_ss\sigma^{s}_{L,\Vert}(T)=0,
\end{align}
\begin{align} \label{f4jxdaroffxtptT}
    \sigma^{o}_{L,\perp}(T)&=\sum_s\sigma^{s}_{L,\perp}(T)\\
    &=-\frac{2e^2}{h}\int^{\infty}_{0}kdk\text{Re}\left\{\frac{\tau^{+-}_{k\uparrow,\perp}}{\hbar}\right\}\frac{A^2(M+Bk^2)}{\mathcal{E}^2_k}\notag\\
    &\times \left[\frac{1}{(e^{\beta(\epsilon_{k+}-\epsilon_{F})} + 1)}-\frac{1}{(e^{\beta(\epsilon_{k-}-\epsilon_{F})} + 1)}\right],\notag 
\end{align}
\begin{align} 
    \sigma^{z}_{L,\perp}(T)=\sum_ss\sigma^{s}_{L,\perp}(T)=0,
\end{align}
\begin{align} \label{f4jxdaroffyttT}
    \sigma^{o}_{H,\Vert}(T)=\sum_s\sigma^{s}_{H,\Vert}(T)=0,
\end{align}
\begin{align} \label{f4jxdaroffyttT}
    \sigma^{z}_{H,\Vert}(T)&=\sum_ss\sigma^{s}_{H,\Vert}(T)\\
    &=-\frac{2e^2}{h}\int^{\infty}_{0}kdk\text{Re}\left\{\frac{\tau^{+-}_{k,\Vert}}{\hbar}\right\}\left[\frac{A^2(M+Bk^2)}{\mathcal{E}^2_k}\right]\notag\\
    &\times \left[\frac{1}{(e^{\beta(\epsilon_{k+}-\epsilon_{F})} + 1)}-\frac{1}{(e^{\beta(\epsilon_{k-}-\epsilon_{F})} + 1)}\right],\notag 
\end{align}
\begin{align} \label{f4jxdaroffytptT}
   \sigma^{o}_{H,\perp}(T)=\sum_s\sigma^{s}_{H,\perp}(T)=0,
\end{align}
\begin{align} \label{f4jxdaroffytptT}
   \sigma^{z}_{H,\perp}&=\sum_ss\sigma^{s}_{H,\perp}\\
   &=\frac{e^2}{h}\int^{\infty}_{0}kdk\text{Im}\left\{\frac{\tau^{+-}_{k\uparrow,\perp}}{\hbar}\right\}\left[\frac{A^2}{\mathcal{E}_k}+\frac{A^2(M+Bk^2)^2}{\mathcal{E}^3_k}\right]\notag\\
   &\times \left[\frac{1}{(e^{\beta(\epsilon_{k+}-\epsilon_{F})} + 1)}-\frac{1}{(e^{\beta(\epsilon_{k-}-\epsilon_{F})} + 1)}\right]. \notag 
\end{align}
At zero temperature, the above conductivities reduce to 
\begin{align} \label{f4jxdaroffxtt}
    \sigma^{o}_{L,\Vert}=\frac{e^2}{h}\int^{\infty}_{0}kdk \text{Im}\left\{\frac{\tau^{+-}_{k,\Vert}}{\hbar}\right\}\left[\frac{A^2}{\mathcal{E}_k}+ \frac{A^2(M+Bk^2)^2}{\mathcal{E}^3_k}\right],
\end{align}
\begin{align} \label{f4jdxdaroffxtt}
    \sigma^{z}_{L,\Vert}=0,
\end{align}
\begin{align} \label{f4jxdaroffxtpt}
    \sigma^{o}_{L,\perp}=\frac{2e^2}{h}\int^{\infty}_{0}kdk\text{Re}\left\{\frac{\tau^{+-}_{k\uparrow,\perp}}{\hbar}\right\}\frac{A^2(M+Bk^2)}{\mathcal{E}^2_k},
\end{align}
\begin{align} 
    \sigma^{z}_{L,\perp}=0,
\end{align}
\begin{align} \label{f4jxdaroffytt}
    \sigma^{o}_{H,\Vert}=0,
\end{align}
\begin{align} \label{f4jxdaroffytt}
    \sigma^{z}_{H,\Vert}=+\frac{2e^2}{h}\int^{\infty}_{0}kdk\text{Re}\left\{\frac{\tau^{+-}_{k,\Vert}}{\hbar}\right\}\left[\frac{A^2(M+Bk^2)}{\mathcal{E}^2_k}\right],
\end{align}
\begin{align} \label{f4jxdaroffytpt}
   \sigma^{o}_{H,\perp}=0,
\end{align}
\begin{align} \label{f4jxdaroffytpt}
   \sigma^{z}_{H,\perp}=-\frac{e^2}{h}\int^{\infty}_{0}kdk\text{Im}\left\{\frac{\tau^{+-}_{k\uparrow,\perp}}{\hbar}\right\}\left[\frac{A^2}{\mathcal{E}_k}+\frac{A^2(M+Bk^2)^2}{\mathcal{E}^3_k}\right].
\end{align}

\section{Derivations of collision integral}
\label{dkfvkdk}

The collision integral, which incorporates the effects of the band structure, is given by, within second-order \textit{Born-Markov approximation}~\cite{zhang2024microscopic,zhang2025open,zhang2026theory}

\begin{widetext}
\begin{align} \label{fagagttgt1da}
   \mathcal{ J}^{\eta_2\eta_1}_{\vec{k}s}(\varrho)&=\frac{2\pi}{V^2} \sum_{\alpha_i} \left\lbrace \mathcal{D}^-(\omega^{\vec{k}_5\eta_5}_{\vec{k}_6\eta_6})\left[e^{-i(\vec{k}-\vec{k}_3)\cdot \vec{r}_{j_1}-i(\vec{k}_5-\vec{k}_6)\cdot \vec{r}_{j_2}} \sigma_{\vec{k}\eta_2,\vec{k}_3\eta_3}^{s} \varrho_{\vec{k}_3s,\vec{k}_5s}^{\eta_3,\eta_5} \sigma_{\vec{k}_5\eta_5,\vec{k}_6\eta_6}^{s}  \bar{\varrho}_{\vec{k}_6s,\vec{k}s}^{\eta_6,\eta_1}\right.\right.\\
    &-\left.e^{-i(\vec{k}_5-\vec{k}_6)\cdot \vec{r}_{j_2}-i(\vec{k}_3-\vec{k})\cdot \vec{r}_{j_1}} \varrho_{\vec{k}s,\vec{k}_5s}^{\eta_2,\eta_5} \sigma_{\vec{k}_5\eta_5,\vec{k}_6\eta_6}^{s}   \bar{\varrho}_{\vec{k}_6s,\vec{k}_3s}^{\eta_6,\eta_3} \sigma_{\vec{k}_3\eta_3,\vec{k}\eta_1}^{s}\right] \notag \\
  &+\mathcal{D}^+(\omega^{\vec{k}_6\eta_6}_{\vec{k}_5\eta_5})\left[e^{-i(\vec{k}_5-\vec{k}_6)\cdot \vec{r}_{j_2}-i(\vec{k}_3-\vec{k})\cdot \vec{r}_{j_1}} \bar{\varrho}_{\vec{k}s,\vec{k}_5s}^{\eta_2,\eta_5}  \sigma_{\vec{k}_5\eta_5,\vec{k}_6\eta_6}^{s}  \varrho_{\vec{k}_6s,\vec{k}_3s}^{\eta_6,\eta_3}   \sigma_{\vec{k}_3\eta_3,\vec{k}\eta_1}^{s}\right.  \notag \\
  &-\left.\left. e^{-i(\vec{k}-\vec{k}_3)\cdot \vec{r}_{j_1}-i(\vec{k}_5-\vec{k}_6)\cdot \vec{r}_{j_2}}  \sigma_{\vec{k}\eta_2,\vec{k}_3\eta_3}^{s}  \bar{\varrho}_{\vec{k}_3s,\vec{k}_5s}^{\eta_3,\eta_5} \sigma_{\vec{k}_5\eta_5,\vec{k}_6\eta_6}^{s}    
   \varrho_{\vec{k}_6s,\vec{k}s}^{\eta_6,\eta_1}\right]\right\},\notag 
\end{align}
\end{widetext}
with $\bar{\varrho}^{\sigma\sigma'}_{\vec{k}s\vec{k}'s}=\delta_{\vec{k}\vec{k}'}\delta_{\sigma\sigma'}-\varrho^{\sigma\sigma'}_{\vec{k}s\vec{k}'s}$, where
\begin{align}
    \mathcal{D}^{+}(\omega) =\frac{1}{2\pi}\int^{+\infty}_{0} d\tau e^{+i\omega\tau-\eta_{\sigma} \tau}\frac{U^2}{\hbar^2},
\end{align}
\begin{align}
    \mathcal{D}^{-}(\omega) =\frac{1}{2\pi}\int^{0}_{-\infty} d\tau e^{+i\omega\tau+\eta_{\sigma} \tau}\frac{U^2}{\hbar^2},
\end{align}
i.e.,
\begin{align} 
     \mathcal{D}^+(\omega) =\frac{1}{2\pi}\frac{U^{2}}{\hbar^2}\frac{0-1}{+i\omega-\eta_{\sigma}}= \frac{1}{2\pi}\frac{U^{2}}{\hbar^2}\left[+\frac{i}{\omega}+\pi\delta(\omega)\right],
\end{align}
\begin{align} 
    \mathcal{D}^-(\omega) =\frac{1}{2\pi}\frac{U^{2}}{\hbar^2} \frac{1-0}{+i\omega+\eta_{\sigma}}=\frac{1}{2\pi}\frac{U^{2}}{\hbar^2}   \left[-\frac{i}{\omega}+\pi\delta(\omega)\right],
\end{align}
which satisfy 
\begin{align} \label{fvdvmdkd}
    \mathcal{D}^-(\omega)=\mathcal{D}^+(-\omega).
\end{align}
The first and third terms (positive sign) of the collision integral \eqref{fagagttgt1da} describe scattering-in processes (gain term), where an electron transitions from a $\vec{k}'$ state  to the $ \vec{k}$ state, thereby increasing the coherence $\delta\varrho^{\bar{\eta}\eta}_{\vec{k}s}$. Conversely, the second and fourth ones  (negative sign) describe scattering-out processes  (loss term), where an electron leaves the $\vec{k}$ state for another $\vec{k}'$ state, leading to a decrease in the  coherence $\delta\varrho^{\bar{\eta}\eta}_{\vec{k}s}$. 

Here, we assume the impurities are randomly distributed, and furthermore, in the dilute limit, we only include the term linear to impurity density $n_i=N_i/V$ by assuming $\vec{r}_{j_2}=\vec{r}_{j_1}$ in Eq.~\eqref{fagagttgt1da}. Here, we use relation of impurity averaging 
\begin{align}
    \frac{1}{V}\sum_{n_1}e^{i(\vec{k}_1-\vec{k}_2)\cdot \vec{r}_{\jmath_1 n_1}}=n_{\text{i}}\delta_{\vec{k}_1,\vec{k}_2},
\end{align}, 
and the collision integral \eqref{fagagttgt1da} reduces to
\begin{align} \label{afdvkdfk}
    \mathcal{ J}^{\eta_1\eta_2}_{\vec{k}s}(\varrho)&=\frac{2\pi n_{\text{i}}}{V} \sum_{\vec{k}'\eta'\eta''} \left\lbrace \left[\mathcal{D}^+(\omega^{\vec{k}\eta_2}_{\vec{k}'\eta'}) \sigma_{\vec{k}\eta_1,\vec{k}'\eta''}^{s} \varrho_{\vec{k}'s}^{\eta''\eta'} \sigma_{\vec{k}'\eta',\vec{k}\eta_2}^{s}\right.\right.\notag\\
    &+\left.\left. \mathcal{D}^+(\omega^{\vec{k}'\eta'}_{\vec{k}\eta_1})  \sigma_{\vec{k}\eta_1,\vec{k}'\eta'}^{s}  \varrho_{\vec{k}'s}^{\eta'\eta''}   \sigma_{\vec{k}'\eta'',\vec{k}\eta_2}^{s} \right] \right.\notag \\
  &-\left.\left[\mathcal{D}^+(\omega^{\vec{k}\eta''}_{\vec{k}'\eta'})  \sigma_{\vec{k}\eta_1,\vec{k}'\eta'}^{s}   \sigma_{\vec{k}'\eta',\vec{k}\eta''}^{s}    
   \varrho_{\vec{k}s}^{\eta''\eta_2}\right.\right. \notag \\
   &+\left.\left. \mathcal{D}^+(\omega^{\vec{k}'\eta'}_{\vec{k}\eta''}) \varrho_{\vec{k}s}^{\eta_1\eta''} \sigma_{\vec{k}\eta'',\vec{k}'\eta'}^{s}   \sigma_{\vec{k}'\eta',\vec{k}\eta_2}^{s}\right]\right\}.  
\end{align}
Linearized to $\vec{E}$, the collision integral \eqref{afdvkdfk} becomes
\begin{align} \label{fvdfvkdkt}
    \mathcal{ J}^{\eta_2\eta_1}_{\vec{k}s}(\delta\varrho)&=\mathcal{ J}^{\eta_2\eta_1}_{\vec{k}s,\text{inv}}(\delta\varrho)+\mathcal{ J}^{\eta_2\eta_1}_{\vec{k}s,\text{del}}(\delta\varrho),
\end{align}
with 
\begin{align} 
   \mathcal{ J}^{\eta_2\eta_1}_{\vec{k}s,\text{inv}}(\delta\varrho)&=i\frac{n_{\text{i}}U^{2}}{\hbar V} \sum_{\vec{k}'\eta'\eta''}  \left\lbrace %
   \frac{\sigma_{\vec{k}\eta_2,\vec{k}'\eta''}^{s} \delta\varrho_{\vec{k}'s}^{\eta''\eta'} \sigma_{\vec{k}'\eta',\vec{k}\eta_1}^{s}}{\epsilon_{\vec{k}\eta_1}-\epsilon_{\vec{k}'\eta'}}  \right. \notag\\
   &+\left.\frac{1}{\epsilon_{\vec{k}'\eta'}-\epsilon_{\vec{k}\eta_2}}  \sigma_{\vec{k}\eta_2,\vec{k}'\eta'}^{s}  \delta\varrho_{\vec{k}'s}^{\eta'\eta''}   \sigma_{\vec{k}'\eta'',\vec{k}\eta_1}^{s} 
    \right.\notag\\
  &\left.- \frac{1}{\epsilon_{\vec{k}\eta''}-\epsilon_{\vec{k}'\eta'}}  \sigma_{\vec{k}\eta_2,\vec{k}'\eta'}^{s}   \sigma_{\vec{k}'\eta',\vec{k}\eta''}^{s}    
   \delta\varrho_{\vec{k}s}^{\eta''\eta_1} \right.\notag \\
   &-\left.\frac{1}{\epsilon_{\vec{k}'\eta'}-\epsilon_{\vec{k}\eta''}} \delta\varrho_{\vec{k}s}^{\eta_2\eta''} \sigma_{\vec{k}\eta'',\vec{k}'\eta'}^{s}   \sigma_{\vec{k}'\eta',\vec{k}\eta_1}^{s} % 
   \right\}, 
\end{align}
\begin{align} \label{fvdfvkdkt}
   \mathcal{ J}^{\eta_2\eta_1}_{\vec{k}s,\text{del}}(\delta\varrho)&=\pi\frac{n_{\text{i}}U^{2}}{\hbar V}\sum_{\vec{k}'\eta'\eta''}  \left\lbrace %
   \delta(\epsilon_{\vec{k}\eta_1}-\epsilon_{\vec{k}'\eta'}) \sigma_{\vec{k}\eta_2,\vec{k}'\eta''}^{s} \delta\varrho_{\vec{k}'s}^{\eta''\eta'} \sigma_{\vec{k}'\eta',\vec{k}\eta_1}^{s} \right.\notag\\
   &+\left.\delta(\epsilon_{\vec{k}'\eta'}-\epsilon_{\vec{k}\eta_2})  \sigma_{\vec{k}\eta_2,\vec{k}'\eta'}^{s}  \delta\varrho_{\vec{k}'s}^{\eta'\eta''}   \sigma_{\vec{k}'\eta'',\vec{k}\eta_1}^{s} \right.\notag\\
  &\left. -\delta(\epsilon_{\vec{k}\eta''}-\epsilon_{\vec{k}'\eta'})  \sigma_{\vec{k}\eta_2,\vec{k}'\eta'}^{s}   \sigma_{\vec{k}'\eta',\vec{k}\eta''}^{s}    
   \delta\varrho_{\vec{k}s}^{\eta''\eta_1}    \right.\notag \\
   &-\left.\delta(\epsilon_{\vec{k}'\eta'}-\epsilon_{\vec{k}\eta''}) \delta\varrho_{\vec{k}s}^{\eta_2\eta''} \sigma_{\vec{k}\eta'',\vec{k}'\eta'}^{s}   \sigma_{\vec{k}'\eta',\vec{k}\eta_1}^{s} 
   \right\}.
\end{align}
Here $\mathcal{ J}^{\eta_2\eta_1}_{\vec{k}s,\text{inv}}(\delta\varrho)$ corresponds to contribution from the principal value of $\mathcal{D}^{\pm}(\omega)$. We will not consider the principal-value contributions $\mathcal{ J}^{\eta_2\eta_1}_{\vec{k}s,\text{inv}}(\delta\varrho)$ to the collision terms further in this work for simplicity,
but note however that they can be incorporated into the general solution using the method outlined in Ref.~\cite{culcer2017interband}. As a result, we attain the collision integral  as follows
\begin{align} \label{fdfkvk}
   \mathcal{ J}^{\eta_2\eta_1}_{\vec{k}s}(\delta\varrho)&=\pi\frac{n_{\text{i}}U^{2}}{\hbar V} \sum_{\vec{k}'\eta'\eta''}  \left\lbrace %
   \delta(\epsilon_{\vec{k}\eta_1}-\epsilon_{\vec{k}'\eta'}) \sigma_{\vec{k}\eta_2,\vec{k}'\eta''}^{s} \varrho_{\vec{k}'s}^{\eta''\eta'} \sigma_{\vec{k}'\eta',\vec{k}\eta_1}^{s}  \right.\notag\\
   &+\left.\delta(\epsilon_{\vec{k}'\eta'}-\epsilon_{\vec{k}\eta_2})  \sigma_{\vec{k}\eta_2,\vec{k}'\eta'}^{s}  \delta\varrho_{\vec{k}'s}^{\eta'\eta''}   \sigma_{\vec{k}'\eta'',\vec{k}\eta_1}^{s} \right.\notag\\
  &\left. -\delta(\epsilon_{\vec{k}\eta''}-\epsilon_{\vec{k}'\eta'})  \sigma_{\vec{k}\eta_2,\vec{k}'\eta'}^{s}   \sigma_{\vec{k}'\eta',\vec{k}\eta''}^{s}    
   \delta\varrho_{\vec{k}s}^{\eta''\eta_1}   \right.\notag\\
   &-\left.\delta(\epsilon_{\vec{k}'\eta'}-\epsilon_{\vec{k}\eta''}) \delta\varrho_{\vec{k}s}^{\eta_2\eta''} \sigma_{\vec{k}\eta'',\vec{k}'\eta'}^{s}   \sigma_{\vec{k}'\eta',\vec{k}\eta_1}^{s} 
   \right\}, 
\end{align}
where $\delta(\epsilon_{\vec{k}\eta}-\epsilon_{\vec{k}'\eta'})$ implies the energy conservation during the scattering events. Notably, both density matrix and collision term satisfy Hermitian, i.e., $\delta\varrho_{\vec{k}s}^{\eta_2\eta_1}=[\delta\varrho_{\vec{k}s}^{\eta_1\eta_2}]^*$ and $\mathcal{ J}^{\eta_2\eta_1}_{\vec{k}s}(\varrho)=[\mathcal{ J}^{\eta_1\eta_2}_{\vec{k}s}(\varrho)]^{*}$.

\subsection{Scattering matrices}

In this subsection, we calculate the scattering matrix, $\sigma_{\vec{k}\eta_1,\vec{k}'\eta_2}^{s}\sigma_{\vec{k}'\eta_3,\vec{k}\eta_4}^{s}$. By means of the eigenfunctions 
\begin{align} 
    \vert \vec{k}s+ \rangle=\begin{bmatrix}
        s\cos\frac{\Theta_{k}}{2}e^{-si\theta_{\vec{k}}}\\
        +\sin\frac{\Theta_{k}}{2}
    \end{bmatrix},
\end{align}
\begin{align} 
    \vert \vec{k}s- \rangle=\begin{bmatrix}
        s\sin\frac{\Theta_{k}}{2}e^{-si\theta_{\vec{k}}}\\
        -\cos\frac{\Theta_{k}}{2}
    \end{bmatrix},
\end{align}
an elementary vector multiplication procedure generates
\begin{align} \label{setaeta1}
    \sigma_{\vec{k}+,\vec{k}'+}^{s}=\cos\frac{\Theta_{k}}{2}\cos\frac{\Theta_{k'}}{2}e^{-si\theta_{\vec{k}'\vec{k}}}+\sin\frac{\Theta_{k}}{2}\sin\frac{\Theta_{k'}}{2},
\end{align}
\begin{align}
    \sigma_{\vec{k}+,\vec{k}'-}^{s}=\cos\frac{\Theta_{k}}{2}\sin\frac{\Theta_{k'}}{2}e^{-si\theta_{\vec{k}'\vec{k}}}-\sin\frac{\Theta_{k}}{2}\cos\frac{\Theta_{k'}}{2},
\end{align}
\begin{align}
    \sigma_{\vec{k}-,\vec{k}'+}^{s}=\sin\frac{\Theta_{k}}{2}\cos\frac{\Theta_{k'}}{2}e^{-si\theta_{\vec{k}'\vec{k}}}-\cos\frac{\Theta_{k}}{2}\sin\frac{\Theta_{k'}}{2},
\end{align}
\begin{align} \label{setaeta4}
    \sigma_{\vec{k}-,\vec{k}'-}^{s}=\sin\frac{\Theta_{k}}{2}\sin\frac{\Theta_{k'}}{2}e^{-si\theta_{\vec{k}'\vec{k}}}+\cos\frac{\Theta_{k}}{2}\cos\frac{\Theta_{k'}}{2},
\end{align}
with $\theta_{\vec{k}'\vec{k}}=\theta_{\vec{k}'}-\theta_{\vec{k}}$.  

Firstly, we are required to derive the following scattering matrix 
\begin{align} \label{fdvkdfk1}
     \vert \sigma_{\vec{k}\eta,\vec{k}'\eta}^{s}\vert^2
     &=\frac{1}{2}\left(1+\cos\Theta_{k}\cos\Theta_{k'}\right)\\
     &+\frac{1}{2}\sin\Theta_{k}\sin\Theta_{k'}\cos\theta_{\vec{k}'\vec{k}}, \notag
\end{align}
\begin{align} \label{fdvkdfk2}
    \vert \sigma_{\vec{k}\eta,\vec{k}'\bar{\eta}}^{s}\vert^2
     &=\frac{1}{2}\left(1-\cos\Theta_{k}\cos\Theta_{k'}\right)\\
     &-\frac{1}{2}\sin\Theta_{k}\sin\Theta_{k'}\cos\theta_{\vec{k}'\vec{k}}, \notag
\end{align}
i.e.
\begin{align} \label{fvavfgdvl}
    \sigma_{\vec{k}\eta,\vec{k}'\eta'}^{s}  \sigma_{\vec{k}'\eta',\vec{k}\eta}^{s}&= \frac{1}{2}\left[1+\eta\eta'\cos\Theta_{k}\cos\Theta_{k'}\right.\notag\\
    &+\left.\eta\eta'\sin\Theta_{k}\sin\Theta_{k'}\cos\theta_{\vec{k}'\vec{k}}\right].
\end{align}
Secondly, we calculate the scattering matrix, $\sigma_{\vec{k}\bar{\eta},\vec{k}'\eta'}^{s}\sigma_{\vec{k}'\eta',\vec{k}\eta}^{s}$. 
First, we can calculate $\sigma_{\vec{k}\bar{\eta},\vec{k}'\eta}^{s}\sigma_{\vec{k}'\eta,\vec{k}\eta}^{s}$ terms as follows 
\begin{align}
    \sigma_{\vec{k}-,\vec{k}'+}^{s}\sigma_{\vec{k}'+,\vec{k}+}^{s}
    &=+\frac{1}{2} \sin\Theta_{k}\cos\Theta_{k'}-is\frac{1}{2}\sin\Theta_{k'}\sin\theta_{\vec{k}'\vec{k}}\notag\\
    &-\frac{1}{2} \sin\Theta_{k'}\cos\Theta_{k}\cos\theta_{\vec{k}'\vec{k}},
\end{align}
\begin{align}
    \sigma_{\vec{k}+,\vec{k}'-}^{s}\sigma_{\vec{k}'-,\vec{k}-}^{s}
    &=-\frac{1}{2} \sin\Theta_{k}\cos\Theta_{k'}-is\frac{1}{2}\sin\Theta_{k'}\sin\theta_{\vec{k}'\vec{k}}\notag\\
    &+\frac{1}{2} \sin\Theta_{k'}\cos\Theta_{k}\cos\theta_{\vec{k}'\vec{k}}.
\end{align}
Then, we work on $\sigma_{\vec{k}\bar{\eta},\vec{k}'\bar{\eta}}^{s}\sigma_{\vec{k}'\bar{\eta},\vec{k}\eta}^{s}$ as follows 
\begin{align}
    \sigma_{\vec{k}-,\vec{k}'-}^{s}&\sigma_{\vec{k}'-,\vec{k}+}^{s}=(\sigma_{\vec{k}+,\vec{k}'-}^{s}\sigma_{\vec{k}'-,\vec{k}-}^{s})^*\\
    &=-\frac{1}{2} \sin\Theta_{k}\cos\Theta_{k'}+\frac{1}{2} \sin\Theta_{k'}\cos\Theta_{k}\cos\theta_{\vec{k}'\vec{k}}\notag\\
    &+si\frac{1}{2} \sin\Theta_{k'}\sin\theta_{\vec{k}'\vec{k}},\notag 
\end{align}
\begin{align}
    \sigma_{\vec{k}+,\vec{k}'+}^{s}&\sigma_{\vec{k}'+,\vec{k}-}^{s}=(\sigma_{\vec{k}-,\vec{k}'+}^{s}\sigma_{\vec{k}'+,\vec{k}+}^{s})^*\\
    &=+\frac{1}{2} \sin\Theta_{k}\cos\Theta_{k'}-\frac{1}{2} \sin\Theta_{k'}\cos\Theta_{k}\cos\theta_{\vec{k}'\vec{k}}\notag\\
    &+si\frac{1}{2} \sin\Theta_{k'}\sin\theta_{\vec{k}'\vec{k}}.\notag 
\end{align}
Thus, we attain
\begin{align} \label{gcgvvjh}
    \sigma_{\vec{k}\bar{\eta},\vec{k}'\eta'}^{s}\sigma_{\vec{k}'\eta',\vec{k}\eta}^{s}&=\frac{\eta'}{2}\left[+ \sin\Theta_{k}\cos\Theta_{k'}-is\eta\sin\Theta_{k'}\sin\theta_{\vec{k}'\vec{k}}\right.\notag \\
    &-\left. \sin\Theta_{k'}\cos\Theta_{k}\cos\theta_{\vec{k}'\vec{k}}\right].
\end{align}
Thirdly, we figure out $\sigma_{\vec{k}\eta,\vec{k}'\eta'}^{s}  \sigma_{\vec{k}'\bar{\eta}',\vec{k}\eta}^{s}$ and  $\sigma_{\vec{k}\bar{\eta},\vec{k}'\eta'}^{s}\sigma_{\vec{k}'\bar{\eta}',\vec{k}\eta}^{s}$. 
The $\sigma_{\vec{k}\eta,\vec{k}'\eta'}^{s}  \sigma_{\vec{k}'\bar{\eta}',\vec{k}\eta}^{s}$ terms can be derived from the conjugation of $\sigma_{\vec{k}\bar{\eta},\vec{k}'\eta'}^{s}\sigma_{\vec{k}'\eta',\vec{k}\eta}^{s}$, which reads  
\begin{align} \label{gbsgbk}
    \sigma_{\vec{k}'\eta',\vec{k}\bar{\eta}}^{s}&\sigma_{\vec{k}\eta,\vec{k}'\eta'}^{s}=\frac{\eta'}{2}\left[ \sin\Theta_{k}\cos\Theta_{k'}+is\eta\sin\Theta_{k'}\sin\theta_{\vec{k}'\vec{k}}\right.\notag \\
    &\left.-\sin\Theta_{k'}\cos\Theta_{k}\cos\theta_{\vec{k}'\vec{k}}\right],
\end{align}
which after exchanging $\eta$ ($\vec{k}$) and $\eta'$ ($\vec{k}'$) becomes 
\begin{align} 
    \sigma_{\vec{k}\eta,\vec{k}'\bar{\eta}'}^{s}&\sigma_{\vec{k}'\eta',\vec{k}\eta}^{s}=\frac{\eta}{2}\left[ \sin\Theta_{k'}\cos\Theta_{k}-is\eta'\sin\Theta_{k}\sin\theta_{\vec{k}'\vec{k}}\right.\notag\\
    &\left.-\sin\Theta_{k}\cos\Theta_{k'}\cos\theta_{\vec{k}'\vec{k}}\right].
\end{align}
Further exchanging $\eta'$ and $\bar{\eta}'$  results in 
\begin{align} \label{gbsgbk}
    \sigma_{\vec{k}\eta,\vec{k}'\eta'}^{s}&\sigma_{\vec{k}'\bar{\eta}',\vec{k}\eta}^{s}=\frac{\eta}{2}\left[+ \sin\Theta_{k'}\cos\Theta_{k}+is\eta'\sin\Theta_{k}\sin\theta_{\vec{k}'\vec{k}}\right.\notag\\
    &\left.-\sin\Theta_{k}\cos\Theta_{k'}\cos\theta_{\vec{k}'\vec{k}}\right].
\end{align}
Besides,  we work on $\sigma_{\vec{k}\bar{\eta},\vec{k}'\eta'}^{s}\sigma_{\vec{k}'\bar{\eta}',\vec{k}\eta}^{s}$ terms. By substitution of Eqs.~(\ref{setaeta1}-\ref{setaeta4}), we attain 
\begin{align} \label{fvfdkdk1}
    \sigma_{\vec{k}+,\vec{k}'+}^{s}\sigma_{\vec{k}'-,\vec{k}-}^{s}
    &=+\frac{1}{2}\sin\Theta_{k}\sin\Theta_{k'}\notag\\
    &+\frac{1}{2}(1+\cos\Theta_{k}\cos\Theta_{k'})\cos\theta_{\vec{k}'\vec{k}}\notag \\
    &-\frac{1}{2}(\cos\Theta_{k}+\cos\Theta_{k'})is\sin\theta_{\vec{k}'\vec{k}},
\end{align}
\begin{align} \label{fvfdkdk2}
    \sigma_{\vec{k}+,\vec{k}'-}^{s}\sigma_{\vec{k}'+,\vec{k}-}^{s}
    &=+\frac{1}{2} \sin\Theta_{k}\sin\Theta_{k'}\notag\\
    &-\frac{1}{2}(1-\cos\Theta_{k}\cos\Theta_{k'})\cos\theta_{\vec{k}'\vec{k}}\notag \\
    &+\frac{1}{2}(\cos\Theta_{k}-\cos\Theta_{k'})is\sin\theta_{\vec{k}'\vec{k}},
\end{align}
whose conjugations are given by
\begin{align} \label{fvfdkdk3}
    \sigma_{\vec{k}-,\vec{k}'-}^{s}\sigma_{\vec{k}'+,\vec{k}+}^{s}
    &=+\frac{1}{2}\sin\Theta_{k}\sin\Theta_{k'}\notag\\
    &+\frac{1}{2}(1+\cos\Theta_{k}\cos\Theta_{k'})\cos\theta_{\vec{k}'\vec{k}}\notag\\
    &+\frac{1}{2}(\cos\Theta_{k}+\cos\Theta_{k'})is\sin\theta_{\vec{k}'\vec{k}},
\end{align}
\begin{align} \label{fvfdkdk4}
    \sigma_{\vec{k}-,\vec{k}'+}^{s}\sigma_{\vec{k}'-,\vec{k}+}^{s}
    &=+\frac{1}{2} \sin\Theta_{k}\sin\Theta_{k'}\notag\\
    &-\frac{1}{2}(1-\cos\Theta_{k}\cos\Theta_{k'})\cos\theta_{\vec{k}'\vec{k}}\notag \\
    &-\frac{1}{2}(\cos\Theta_{k}-\cos\Theta_{k'})is\sin\theta_{\vec{k}'\vec{k}}. 
\end{align}
Combining Eqs. (\ref{fvfdkdk1}-\ref{fvfdkdk4}) generates
\begin{align} \label{fvdkvdkfkv1}
    \sigma_{\vec{k}\bar{\eta},\vec{k}'\bar{\eta}}^{s}  \sigma_{\vec{k}'\eta,\vec{k}\eta}^{s}&=+\frac{1}{2}\sin\Theta_{k}\sin\Theta_{k'}\notag\\
    &+\frac{1}{2}(1+\cos\Theta_{k}\cos\Theta_{k'})\cos(\theta_{\vec{k}'\vec{k}})\notag\\
    &+s\eta\frac{i}{2}(\cos\Theta_{k}+\cos\Theta_{k'})\sin(\theta_{\vec{k}'\vec{k}}),
\end{align}
\begin{align} \label{fvdkvdkfkv2}
    \sigma_{\vec{k}\bar{\eta},\vec{k}'\eta}^{s}\sigma_{\vec{k}'\bar{\eta},\vec{k}\eta}^{s}&=+\frac{1}{2} \sin\Theta_{k}\sin\Theta_{k'}\notag\\
    &-\frac{1}{2}(1-\cos\Theta_{k}\cos\Theta_{k'})\cos(\theta_{\vec{k}'\vec{k}})\notag\\
    &-s\eta\frac{i}{2}(\cos\Theta_{k}-\cos\Theta_{k'})\sin(\theta_{\vec{k}'\vec{k}}). 
\end{align}
Finally the scattering matrices $\sigma_{\vec{k}\eta_1,\vec{k}'\eta_2}^{s}\sigma_{\vec{k}'\eta_3,\vec{k}\eta_4}^{s}$ are separated into following five terms 
\begin{widetext}
\begin{align} \label{fvavfgdvl1}
    \sigma_{\vec{k}\eta,\vec{k}'\eta'}^{s}  \sigma_{\vec{k}'\eta',\vec{k}\eta}^{s}= \frac{1}{2}\left[1+\eta\eta'\cos\Theta_{k}\cos\Theta_{k'}+\eta\eta'\sin\Theta_{k}\sin\Theta_{k'}\cos\theta_{\vec{k}'\vec{k}}\right].
\end{align}
\begin{align} \label{fvavfgdvl2}
    \sigma_{\vec{k}\bar{\eta},\vec{k}'\eta'}^{s}\sigma_{\vec{k}'\eta',\vec{k}\eta}^{s}=\frac{\eta'}{2}\left[+ \sin\Theta_{k}\cos\Theta_{k'}-\sin\Theta_{k'}\cos\Theta_{k}\cos\theta_{\vec{k}'\vec{k}}-is\eta\sin\Theta_{k'}\sin\theta_{\vec{k}'\vec{k}}\right].
\end{align}
\begin{align} \label{fvavfgdvl3}
    \sigma_{\vec{k}\eta,\vec{k}'\eta'}^{s}\sigma_{\vec{k}'\bar{\eta}',\vec{k}\eta}^{s}=\frac{\eta}{2}\left[+ \sin\Theta_{k'}\cos\Theta_{k}-\sin\Theta_{k}\cos\Theta_{k'}\cos\theta_{\vec{k}'\vec{k}}+is\eta'\sin\Theta_{k}\sin\theta_{\vec{k}'\vec{k}}\right].
\end{align}
\begin{align} \label{fvavfgdvl4}
    \sigma_{\vec{k}\bar{\eta},\vec{k}'\bar{\eta}}^{s}  \sigma_{\vec{k}'\eta,\vec{k}\eta}^{s}=+\frac{1}{2}\sin\Theta_{k}\sin\Theta_{k'}+\frac{1}{2}(1+\cos\Theta_{k}\cos\Theta_{k'})\cos(\theta_{\vec{k}'\vec{k}})+s\eta\frac{i}{2}(\cos\Theta_{k}+\cos\Theta_{k'})\sin(\theta_{\vec{k}'\vec{k}}),
\end{align}
\begin{align} \label{fvavfgdvl5}
    \sigma_{\vec{k}\bar{\eta},\vec{k}'\eta}^{s}\sigma_{\vec{k}'\bar{\eta},\vec{k}\eta}^{s}=+\frac{1}{2} \sin\Theta_{k}\sin\Theta_{k'}-\frac{1}{2}(1-\cos\Theta_{k}\cos\Theta_{k'})\cos(\theta_{\vec{k}'\vec{k}})-s\eta\frac{i}{2}(\cos\Theta_{k}-\cos\Theta_{k'})\sin(\theta_{\vec{k}'\vec{k}}). 
\end{align}
\end{widetext}

\subsection{Ordinary collision integral}
First, we recover the standard Fermi Golden rule, based on the intraband collision integral, $ \mathcal{ J}^{\eta\eta}_{\vec{k}}(\delta\varrho)$. To this end, we exclude the interband correlation of the density matrix by assuming the following diagonal approximation 
\begin{align} \label{gbgfb}
    \varrho_{\vec{k}s}^{\eta_1\eta_2}\simeq \delta_{\eta_1\eta_2}\varrho_{\vec{k}s\eta_1}. 
\end{align}
Then, the intraband collision term [i.e., $\eta_1=\eta$ and $\eta_2=\eta$ terms of Eq. \eqref{fdfkvk}] reduces to
\begin{align} 
   \mathcal{ J}^{\eta\eta}_{\vec{k}s}(\delta\varrho)&=\pi\frac{n_{\text{i}}U^{2}}{\hbar V}\sum_{\vec{k}'\eta'}  \left\lbrace %
   \delta(\epsilon_{\vec{k}\eta}-\epsilon_{\vec{k}'\eta'}) \sigma_{\vec{k}\eta,\vec{k}'\eta'}^{s} \delta\varrho_{\vec{k}'s\eta'} \sigma_{\vec{k}'\eta',\vec{k}\eta}^{s} \right.\notag\\
   &\left.+\delta(\epsilon_{\vec{k}'\eta'}-\epsilon_{\vec{k}\eta})  \sigma_{\vec{k}\eta,\vec{k}'\eta'}^{s}  \delta\varrho_{\vec{k}'s\eta'}   \sigma_{\vec{k}'\eta',\vec{k}\eta}^{s} \right.\\
  &\left. -\delta(\epsilon_{\vec{k}\eta}-\epsilon_{\vec{k}'\eta'})  \sigma_{\vec{k}\eta,\vec{k}'\eta'}^{s}   \sigma_{\vec{k}'\eta',\vec{k}\eta}^{s}    
   \delta\varrho_{\vec{k}s\eta}   \right. \notag\\
   &\left.-\delta(\epsilon_{\vec{k}'\eta'}-\epsilon_{\vec{k}\eta}) \delta\varrho_{\vec{k}s\eta} \sigma_{\vec{k}\eta,\vec{k}'\eta'}^{s}   \sigma_{\vec{k}'\eta',\vec{k}\eta}^{s} 
   \right\},\notag
\end{align}
i.e.,
\begin{align} \label{fvnafkvmn}
  \mathcal{ J}^{\eta\eta}_{\vec{k}s}(\varrho)&=\frac{1}{V}\sum_{\vec{k}'\eta'} W_{\eta,\eta'}(\vec{k},\vec{k}') (\delta\varrho_{\vec{k}'s\eta'} - \delta\varrho_{\vec{k}s\eta} ) .
\end{align}
Within the diagonal approximation  \eqref{gbgfb}, the collision contribution to the quantum kinetic equation can be
viewed as a matrix operator acting on the density matrix, which defines the following scattering matrix 
\begin{align} \label{fsvvls1}
    W_{\eta,\eta'}(\vec{k},\vec{k}')=2\pi n_{\text{i}} \frac{U^{2}}{\hbar}  \vert \sigma_{\vec{k}\eta,\vec{k}'\eta'}^{s}\vert^2 \delta(\epsilon_{\vec{k}\eta}-\epsilon_{\vec{k}'\eta'}).
\end{align}
Therefore, we recover the standard Fermi Golden rule. The scattering matrix, $\vert \sigma_{\vec{k}\eta,\vec{k}'\eta'}^{s}\vert^2$, can be calculated from the eigenfunctions, as shown in previous subsection [see Eq. \eqref{fvavfgdvl}]. Then, the intraband collision term  \eqref{fvnafkvmn} can be divided into two parts as follows 
\begin{align}  \label{fvfdvkd}
     \mathcal{ J}^{\eta\eta}_{\vec{k}s}(\delta\varrho)= \mathcal{ J}^{\eta\eta}_{\vec{k}s,\text{ra}}(\delta\varrho)+ \mathcal{ J}^{\eta\eta}_{\vec{k}s,\text{er}}(\delta\varrho),
\end{align}
with
\begin{widetext}
\begin{align} \label{ffdfdvdfk1}
  \mathcal{ J}^{\eta\eta}_{\vec{k}s,\text{ra}}(\delta\varrho)&= 2\pi \frac{n_{\text{i}}U^{2}}{\hbar}\frac{1}{V}\sum_{\vec{k}'}\frac{1}{2}\left[1+\cos\Theta_{k}\cos\Theta_{k'}+\sin\Theta_{k}\sin\Theta_{k'}\cos\theta_{\vec{k}'\vec{k}}\right]\delta(\epsilon_{k\eta}-\epsilon_{k'\eta})  (\delta\varrho_{\vec{k}'s\eta}-\delta\varrho_{\vec{k}s\eta}),
\end{align}
\begin{align} \label{ffdfdvdfk2}
  \mathcal{ J}^{\eta\eta}_{\vec{k}s,\text{er}}(\delta\varrho)&=2\pi \frac{n_{\text{i}}U^{2}}{\hbar}\frac{1}{V}\sum_{\vec{k}'}\frac{1}{2}\left[1-\cos\Theta_{k}\cos\Theta_{k'}-\sin\Theta_{k}\sin\Theta_{k'}\cos\theta_{\vec{k}'\vec{k}}\right]\delta(\epsilon_{k\eta}-\epsilon_{k'\bar{\eta}})  (\delta\varrho_{\vec{k}'s\bar{\eta}}-\delta\varrho_{\vec{k}s\eta}).
\end{align}
\end{widetext}
Here, $\mathcal{ J}^{\eta\eta}_{\vec{k}s,\text{ra}}$ describes the contribution from the intra-band scattering events, while $\mathcal{ J}^{\eta\eta}_{\vec{k}s,\text{er}}$ describes the contribution from the interband scattering events. Hereafter, we omit the interband ones because the interband scattering event is energy unfavorable. As a result, we attain the intraband collision integral as follows

\begin{align} \label{ffddfdvdfk1}
  \mathcal{ J}^{\eta\eta}_{\vec{k}s}(\delta\varrho)&\simeq 2\pi \frac{n_{\text{i}}U^{2}}{\hbar}\frac{1}{V}\sum_{\vec{k}'}\frac{1}{2}\left[1+\cos^2\Theta_{k}+\sin^2\Theta_{k}\cos\theta_{\vec{k}'\vec{k}}\right]\notag \\
  &\times \delta(\epsilon_{k\eta}-\epsilon_{k'\eta})  (\delta\varrho_{\vec{k}'s\eta}-\delta\varrho_{\vec{k}s\eta}).
\end{align}

Next, we derive the above collision integral \eqref{ffddfdvdfk1} by using the following ansatz 
\begin{align} 
    \delta\varrho_{\vec{k}s\eta}\simeq -e\tau f'_{k\eta}\vec{v}_{\vec{k}s\eta}\cdot \vec{E},
\end{align}
which satisfy
\begin{align}
    \int^{2\pi}_{0}d\theta_{\vec{k}'}\delta\varrho_{\vec{k}'s\eta} =0,
\end{align}
\begin{align}
    \int^{2\pi}_{0}d\theta_{\vec{k}'}\cos\theta_{\vec{k}'\vec{k}}\delta\varrho_{\vec{k}'s\eta} =\pi \delta\varrho_{\vec{k}s\eta}.
\end{align}
Then, we replace the sum over $\vec{k}'$ with integral, i.e., 
\begin{align} \label{fvfvkgmk}
    \frac{1}{V}\sum_{\vec{k}'}\rightarrow \frac{1}{(2\pi)^2}\int^{\infty}_0 k' dk' \int^{2\pi}_0 d\theta_{\vec{k}'} \rightarrow \frac{1}{2\pi}\int d\epsilon \nu(\epsilon) \int^{2\pi}_0 d\theta_{\vec{k}'}.
\end{align}
The density of state for each energy band is defined by 
\begin{align}
    \nu_{\eta}(\epsilon)&=\frac{1}{V}\sum_{\vec{k}'}\delta(\epsilon-\epsilon_{\vec{k}'\eta})=\int \frac{d \vec{k}'}{(2\pi)^2} \delta(\epsilon-\epsilon_{k'\eta})\notag \\
    &=\frac{1}{2\pi}\int^{\infty}_0 k' dk' \delta(\epsilon-\epsilon_{k'\eta}).
\end{align}
Noting that 
\begin{align} \label{deltaderiv}
    \delta(g(x))=\sum_{i}\frac{\delta(x-x_i)}{\vert g'(x_i) \vert},
\end{align}
with $x_i$  being the $i$th real root of $g(x)$, we attain 
\begin{align}
    \delta(\epsilon-\epsilon_{k'\eta})=\left.\frac{\delta(k'-k)}{\hbar \vert v_{k\eta} \vert}\right\vert_{\epsilon_{k\eta}=\epsilon},
\end{align}
with
\begin{align}
    v_{k\eta}=\frac{1}{\hbar} \partial_k \epsilon_{k'\eta}=\eta\frac{1}{\hbar} \partial_k\mathcal{E}_{k}=\eta k\frac{A^2-2BE^-_k}{\hbar \mathcal{E}_{k}} .
\end{align}
Then, we reach 
\begin{align}
    \nu_{\eta}(\epsilon)=\frac{1}{2\pi} \left.\frac{k}{\hbar \vert v_{k\eta} \vert}\right\vert_{\epsilon_{k\eta}=\epsilon}=\frac{1}{2\pi} \left.\frac{\mathcal{E}_{k}}{  \vert A^2-2BE^-_k \vert }\right\vert_{\epsilon_{k\eta}=\epsilon},
\end{align}
i.e.,
\begin{align}
    \nu_{\eta}(\epsilon_{k\eta})=\frac{\mathcal{E}_k}{2\pi\vert A^2-2BE^-_k\vert }.
\end{align}
By substitution of Eq.~\eqref{fvfvkgmk},
equation~\eqref{ffddfdvdfk1} becomes
\begin{align} \label{frfdfdvdfk1}
  \mathcal{ J}^{\eta\eta}_{\vec{k}s}(\delta\varrho)&= -\pi \frac{n_{\text{i}}\nu_{\eta}(\epsilon_{k\eta})U^{2}}{\hbar} (1+\cos^2\Theta_{k})  \delta\varrho_{\vec{k}s\eta}\\
  &+ \frac{\pi}{2}\frac{n_{\text{i}}\nu_{\eta}(\epsilon_{k\eta})U^{2}}{\hbar} \sin^2\Theta_{k}  \delta\varrho_{\vec{k}s\eta}.\notag 
\end{align}
Therefore, we attain the intraband collision integral 
\begin{align} \label{fvdvdkfk}
    \mathcal{ J}^{\eta\eta}_{\vec{k}s}(\delta\varrho)&=-\frac{1}{\tau^{\text{d}}_{k\eta}}\delta\varrho_{\vec{k}s\eta},
\end{align}
with
\begin{align}
    \frac{\tau^0_{k\eta}}{\tau^{\text{d}}_{k\eta}}=1-\frac{3}{4}\sin^2\Theta_{k},
\end{align}
\begin{align}
    \frac{1}{\tau^0_{k\eta}}=2\pi \frac{n_{i}\nu_{\eta}(\epsilon_{k\eta})U^{2}}{\hbar} .
\end{align}
Therefore, we recover the collision integral under the standard momentum relaxation approximation.

\subsection{Anomalous collision integral}

In this subsubsection, we go beyond the diagonal approximation of density matrix \eqref{gbgfb} and focus on the interband collision integral, $\mathcal{ J}^{\bar{\eta}\eta}_{\vec{k}s}(\delta\varrho)$. 
The $\eta_1=\eta$ and $\eta_2=\bar{\eta}$ term of Eq. \eqref{fdfkvk} becomes
\begin{align} \label{fvdfvkdktdfa}
    \mathcal{ J}^{\bar{\eta}\eta}_{\vec{k}s}(\delta\varrho)&=\mathcal{ J}^{\bar{\eta}\eta}_{\vec{k}s,\text{dia}}(\delta\varrho)+\mathcal{ J}^{\bar{\eta}\eta}_{\vec{k}s,\text{off}}(\delta\varrho),
\end{align}
with 
\begin{align} \label{fvdfvkdfsa}
     \mathcal{ J}^{\bar{\eta}\eta}_{\vec{k}s,\text{dia}}(\delta\varrho)&= \frac{\pi \hbar D_n}{ V}\sum_{\vec{k}'\eta'}\sigma_{\vec{k}\bar{\eta},\vec{k'}\eta'}^{s}   \sigma_{\vec{k'}\eta',\vec{k}\eta}^{s} \left\{\delta(\epsilon_{\vec{k}\eta}-\epsilon_{\vec{k}'\eta'})(\varrho_{\vec{k}'s\eta'}-\varrho_{\vec{k}s\eta}    )\right.\notag\\
     &\left.+\delta(\epsilon_{\vec{k}'\eta'}-\epsilon_{\vec{k}\bar{\eta}})(\varrho_{\vec{k}'s\eta'} -\varrho_{\vec{k}s\bar{\eta}} )\right\},
\end{align}
\begin{align} \label{fdfvvolqfami}
  \mathcal{ J}^{\bar{\eta}\eta}_{\vec{k}s,\text{off}}(\delta\varrho)&=+\frac{\pi \hbar D_n}{V} \sum_{\vec{k}'\eta'} \left\lbrace 
   \delta(\epsilon_{\vec{k}'\bar{\eta}'}-\epsilon_{\vec{k}\eta}) \sigma_{\vec{k}\bar{\eta},\vec{k}'\eta'}^{s} \delta\varrho_{\vec{k}'s}^{\eta'\bar{\eta}'} \sigma_{\vec{k}'\bar{\eta}',\vec{k}\eta}^{s}   \right.\notag\\
   &\left.+\delta(\epsilon_{\vec{k}\bar{\eta}}-\epsilon_{\vec{k}'\eta'})  \sigma_{\vec{k}\bar{\eta},\vec{k}'\eta'}^{s}   \delta\varrho_{\vec{k}'s}^{\eta'\bar{\eta}'} \sigma_{\vec{k}'\bar{\eta}',\vec{k}\eta}^{s}    
   \right\}\\
   &-\frac{\pi \hbar D_n}{V} \sum_{\vec{k}'\eta'} \left\lbrace %
   \delta(\epsilon_{\vec{k}'\eta'}-\epsilon_{\vec{k}\bar{\eta}}) \sigma_{\vec{k}\bar{\eta},\vec{k}'\eta'}^{s}  \sigma_{\vec{k}'\eta',\vec{k}\bar{\eta}}^{s}  \delta\varrho_{\vec{k}s}^{\bar{\eta}\eta}\right.\notag\\
   &\left.+\delta(\epsilon_{\vec{k}\eta}-\epsilon_{\vec{k}'\eta'}) \delta\varrho_{\vec{k}s}^{\bar{\eta}\eta} \sigma_{\vec{k}\eta,\vec{k}'\eta'}^{s}   \sigma_{\vec{k}'\eta',\vec{k}\eta}^{s} %   
   \right\} .\notag 
\end{align} 
with
\begin{align}
    D_n=\frac{ n_{\text{i}}U^{2}}{\hbar^2}.
\end{align}
Here, $\mathcal{ J}^{\bar{\eta}\eta}_{\vec{k}s,\text{dia}}(\delta\varrho)$ describes the contribution from the diagonal density matrix, while $\mathcal{ J}^{\bar{\eta}\eta}_{\vec{k}s,\text{off}}(\delta\varrho)$ describes the contribution from the off-diagonal density matrix. Hereafter, we work on quantum decoherence--the decay of the off-diagonal component of density matrix, and thus we focus on collision integrals $\mathcal{ J}^{\bar{\eta}\eta}_{\vec{k}s,\text{off}}(\delta\varrho)$, which can be divided into two parts as follows   
\begin{align}
    \mathcal{ J}^{\bar{\eta}\eta}_{\vec{k}s,\text{off}}(\delta\varrho)=\mathcal{ J}^{\bar{\eta}\eta}_{\vec{k}s,\text{off},\text{ra}}(\delta\varrho)+\mathcal{ J}^{\bar{\eta}\eta}_{\vec{k}s,\text{off},\text{er}}(\delta\varrho),
\end{align}
with
\begin{align} 
   \mathcal{ J}^{\bar{\eta}\eta}_{\vec{k}s,\text{off},\text{ra}}(\delta\varrho)&=+\frac{\pi \hbar D_n}{V} \sum_{\vec{k}'} \left\lbrace 
   +\delta(\epsilon_{k'\eta}-\epsilon_{k\eta}) \sigma_{\vec{k}\bar{\eta},\vec{k}'\bar{\eta}}^{s} \delta\varrho_{\vec{k}'s}^{\bar{\eta}\eta} \sigma_{\vec{k}'\eta,\vec{k}\eta}^{s} \right.\notag\\
   &\left.-\delta(\epsilon_{\vec{k}\eta}-\epsilon_{\vec{k}'\eta}) \delta\varrho_{\vec{k}s}^{\bar{\eta}\eta} \sigma_{\vec{k}\eta,\vec{k}'\eta}^{s}   \sigma_{\vec{k}'\eta,\vec{k}\eta}^{s}    
   \right\}\\
   &+\frac{\pi \hbar D_n}{V} \sum_{\vec{k}'} \left\lbrace %
   -\delta(\epsilon_{k'\bar{\eta}}-\epsilon_{k\bar{\eta}}) \sigma_{\vec{k}\bar{\eta},\vec{k}'\bar{\eta}}^{s}  \sigma_{\vec{k}'\bar{\eta},\vec{k}\bar{\eta}}^{s}  \delta\varrho_{\vec{k}s}^{\bar{\eta}\eta} \right.\notag\\
   &+\left.\delta(\epsilon_{\vec{k}\bar{\eta}}-\epsilon_{\vec{k}'\bar{\eta}})  \sigma_{\vec{k}\bar{\eta},\vec{k}'\bar{\eta}}^{s}   \delta\varrho_{\vec{k}'s}^{\bar{\eta}\eta} \sigma_{\vec{k}'\eta,\vec{k}\eta}^{s} %   
   \right\} ,\notag 
\end{align}
\begin{align} \label{efdfvrvolqplf2}
   \mathcal{ J}^{\bar{\eta}\eta}_{\vec{k}s,\text{off},\text{er}}(\delta\varrho)=&+\frac{2\pi \hbar D_n}{V} \sum_{\vec{k}'} \left\lbrace 
   \delta(\epsilon_{k'\bar{\eta}}-\epsilon_{k\eta}) \sigma_{\vec{k}\bar{\eta},\vec{k}'\eta}^{s} \delta\varrho_{\vec{k}'s}^{\eta\bar{\eta}} \sigma_{\vec{k}'\bar{\eta},\vec{k}\eta}^{s}  \right. \notag\\
   &\left.-\delta(\epsilon_{\vec{k}\eta}-\epsilon_{\vec{k}'\bar{\eta}}) \delta\varrho_{\vec{k}s}^{\bar{\eta}\eta} \sigma_{\vec{k}\eta,\vec{k}'\bar{\eta}}^{s}   \sigma_{\vec{k}'\bar{\eta},\vec{k}\eta}^{s}    
   \right\}\\
   &+\frac{2\pi \hbar D_n}{V} \sum_{\vec{k}'} \left\lbrace %
   -\delta(\epsilon_{k'\eta}-\epsilon_{k\bar{\eta}}) \sigma_{\vec{k}\bar{\eta},\vec{k}'\eta}^{s}  \sigma_{\vec{k}'\eta,\vec{k}\bar{\eta}}^{s}  \delta\varrho_{\vec{k}s}^{\bar{\eta}\eta} \right.\notag\\
   &\left.+\delta(\epsilon_{\vec{k}\bar{\eta}}-\epsilon_{\vec{k}'\eta})  \sigma_{\vec{k}\bar{\eta},\vec{k}'\eta}^{s}   \delta\varrho_{\vec{k}'s}^{\eta\bar{\eta}} \sigma_{\vec{k}'\bar{\eta},\vec{k}\eta}^{s} %   
   \right\} .\notag 
\end{align}
Here, $\mathcal{ J}^{\bar{\eta}\eta}_{\vec{k}s,\text{off},\text{ra}}$ describes the contribution from the intra-band scattering events, while $\mathcal{ J}^{\bar{\eta}\eta}_{\vec{k}s,\text{off},\text{er}}$ describes the contribution from the inter-band scattering events. Hereafter, we omit the inter-band contribution because the interband scattering event is energy unfavorable. 
As a result, we reach 
\begin{align} \label{efdfvrvolqplf1}
   \mathcal{ J}^{\bar{\eta}\eta}_{\vec{k}s,\text{off}}(\delta\varrho)&\simeq+\frac{\pi \hbar D_n}{V} \sum_{\vec{k}'} \left\lbrace 
   +\delta(\epsilon_{k'\eta}-\epsilon_{k\eta}) \sigma_{\vec{k}\bar{\eta},\vec{k}'\bar{\eta}}^{s} \delta\varrho_{\vec{k}'s}^{\bar{\eta}\eta} \sigma_{\vec{k}'\eta,\vec{k}\eta}^{s}  \right.\notag\\
   &\left.-\delta(\epsilon_{\vec{k}\eta}-\epsilon_{\vec{k}'\eta}) \delta\varrho_{\vec{k}}^{\bar{\eta}\eta} \sigma_{\vec{k}\eta,\vec{k}'\eta}^{s}   \sigma_{\vec{k}'\eta,\vec{k}\eta}^{s}    
   \right\}\\
   &+\frac{\pi \hbar D_n}{V} \sum_{\vec{k}'} \left\lbrace %
   -\delta(\epsilon_{k'\bar{\eta}}-\epsilon_{k\bar{\eta}}) \sigma_{\vec{k}\bar{\eta},\vec{k}'\bar{\eta}}^{s}  \sigma_{\vec{k}'\bar{\eta},\vec{k}\bar{\eta}}^{s}  \delta\varrho_{\vec{k}s}^{\bar{\eta}\eta} \right.\notag\\
   &\left.+\delta(\epsilon_{\vec{k}\bar{\eta}}-\epsilon_{\vec{k}'\bar{\eta}})  \sigma_{\vec{k}\bar{\eta},\vec{k}'\bar{\eta}}^{s}   \delta\varrho_{\vec{k}'s}^{\bar{\eta}\eta} \sigma_{\vec{k}'\eta,\vec{k}\eta}^{s} %   
   \right\} ,\notag 
\end{align}

Next we derive the above collision term.  To this end, we use the following off-diagonal density matrix 
\begin{align}
    \delta\varrho^{\bar{\eta}\eta}_{\vec{k}s}=\delta\varrho^{\bar{\eta}\eta}_{\vec{k}s,\Vert}+\delta\varrho^{\bar{\eta}\eta}_{\vec{k}s,\perp}, 
\end{align}
with
\begin{align}
    \delta\varrho^{\bar{\eta}\eta}_{\vec{k}s,\Vert}=-e\frac{\tau^{\bar{\eta}\eta}_{ks,\Vert}}{\hbar}(f_{k\bar{\eta}}- f_{k\eta})(\mathcal{R}^{x,\bar{\eta}\eta}_{\vec{k}s} E^x+\mathcal{R}^{y,\bar{\eta}\eta}_{\vec{k}s} E^y),
\end{align}
\begin{align}
\delta\varrho^{\bar{\eta}\eta}_{\vec{k}s,\perp} \simeq -e\frac{\tau^{\bar{\eta}\eta}_{\vec{k}s,\perp}}{\hbar}(f_{k\bar{\eta}}- f_{k\eta})(\mathcal{R}^{x,\bar{\eta}\eta}_{\vec{k}s} E^y-\mathcal{R}^{y,\bar{\eta}\eta}_{\vec{k}s} E^x), 
\end{align}
where
\begin{align}
    \mathcal{R}^{x,\eta\bar{\eta}}_{\vec{k}s}=-s\frac{A}{2\mathcal{E}_{k}}\sin \theta_{\vec{k}}+i\eta\frac{A(M+Bk^2)}{2\mathcal{E}^2_k}\cos\theta_{\vec{k}},
\end{align}
\begin{align}
    \mathcal{R}^{y,\eta\bar{\eta}}_{\vec{k}s}=+s\frac{A}{2\mathcal{E}_{k}}\cos \theta_{\vec{k}}+i\eta\frac{A(M+Bk^2)}{2\mathcal{E}^2_k}\sin\theta_{\vec{k}}.
\end{align}
Notably, $\vec{\mathcal{R}}^{\eta\bar{\eta}}_{\vec{k}s}$ satisfies the following identities
\begin{align} \label{fvdvk1}
    \int_{0}^{2\pi} d\theta_{\vec{k}'} \cos\theta_{\vec{k}'\vec{k}} \tilde{\mathcal{R}}^{x,\eta\bar{\eta}}_{\vec{k}'s}= +\pi\tilde{\mathcal{R}}^{x,\eta\bar{\eta}}_{\vec{k}s},
\end{align}
\begin{align}
    \int_{0}^{2\pi} d\theta_{\vec{k}'} \cos\theta_{\vec{k}'\vec{k}} \tilde{\mathcal{R}}^{y,\eta\bar{\eta}}_{\vec{k}'s}=+\pi\tilde{\mathcal{R}}^{y,\eta\bar{\eta}}_{\vec{k}s},
\end{align}
\begin{align}
    \int_{0}^{2\pi} d\theta_{\vec{k}'} \sin\theta_{\vec{k}'\vec{k}} \tilde{\mathcal{R}}^{x,\eta\bar{\eta}}_{\vec{k}'s}= -\pi\tilde{\mathcal{R}}^{y,\eta\bar{\eta}}_{\vec{k}s},
\end{align}
\begin{align} \label{fvdvk4}
    \int_{0}^{2\pi} d\theta_{\vec{k}'} \sin\theta_{\vec{k}'\vec{k}} \tilde{\mathcal{R}}^{y,\eta\bar{\eta}}_{\vec{k}'s}= +\pi\tilde{\mathcal{R}}^{x,\eta\bar{\eta}}_{\vec{k}s}.
\end{align}
Then, we attain
\begin{align} \label{rfvdvk1}
    \int_{0}^{2\pi} d\theta_{\vec{k}'} \cos\theta_{\vec{k}'\vec{k}} \delta\varrho^{\bar{\eta}\eta}_{\vec{k}'s,\Vert}= +\pi\delta\varrho^{\bar{\eta}\eta}_{\vec{k}s,\Vert},
\end{align}
\begin{align}
    \int_{0}^{2\pi} d\theta_{\vec{k}'} \cos\theta_{\vec{k}'\vec{k}} \delta\varrho^{\bar{\eta}\eta}_{\vec{k}'s,\perp}=+\pi\delta\varrho^{\bar{\eta}\eta}_{\vec{k}s,\perp},
\end{align}
\begin{align}
    \int_{0}^{2\pi} d\theta_{\vec{k}'} \sin\theta_{\vec{k}'\vec{k}} \delta\varrho^{\bar{\eta}\eta}_{\vec{k}'s,\Vert}= +\pi \frac{\tau^{\bar{\eta}\eta}_{\vec{k}s,\Vert}}{\tau^{\bar{\eta}\eta}_{\vec{k}s,\perp}}\delta\varrho^{\bar{\eta}\eta}_{\vec{k}s,\perp},
\end{align}
\begin{align} \label{rfvdvk4}
    \int_{0}^{2\pi} d\theta_{\vec{k}'} \sin\theta_{\vec{k}'\vec{k}} \delta\varrho^{\bar{\eta}\eta}_{\vec{k}'s,\perp}= -\pi\frac{\tau^{\bar{\eta}\eta}_{\vec{k}s,\perp}}{\tau^{\bar{\eta}\eta}_{\vec{k}s,\Vert}}\delta\varrho^{\bar{\eta}\eta}_{\vec{k}s,\Vert}.
\end{align}

By means of the scattering matrices \eqref{fvavfgdvl4} and \eqref{fvdkvdkfkv1}, 
the collision term \eqref{efdfvrvolqplf1} becomes 
\begin{align} \label{efdfvrvolqpavlf1}
   \mathcal{ J}^{\bar{\eta}\eta}_{\vec{k}s,\text{off},\text{ra}}(\delta\varrho)&=\mathcal{ J}^{\bar{\eta}\eta}_{\vec{k}s,\text{off},\text{ra}-1}+\mathcal{ J}^{\bar{\eta}\eta}_{\vec{k}s,\text{off},\text{ra}-2}\\
   &+\mathcal{ J}^{\bar{\eta}\eta}_{\vec{k}s,\text{off},\text{ra}-3}+\mathcal{ J}^{\bar{\eta}\eta}_{\vec{k}s,\text{off},\text{ra}-4},\notag 
\end{align}
with
\begin{widetext}
\begin{align}
     \mathcal{ J}^{\bar{\eta}\eta}_{\vec{k}s,\text{off},\text{ra}-1}&=\frac{\pi \hbar D_n}{V}\frac{V}{(2\pi)^2}\int_{0}^{\infty} k'dk' \int_{0}^{2\pi} d\theta_{\vec{k}'}\delta(\epsilon_{k'\eta}-\epsilon_{k\eta}) \sigma_{\vec{k}\bar{\eta},\vec{k}'\bar{\eta}}^{s} \sigma_{\vec{k}'\eta,\vec{k}\eta}^{s}  \delta\varrho_{\vec{k}'s}^{\bar{\eta}\eta}\\
     &= \frac{\pi D_n}{4\pi^2} \frac{k}{ \left\vert v^{0}_{k\eta} \right\vert} \int_{0}^{2\pi} d\theta_{\vec{k}'}\frac{1}{2}[(1+\cos\Theta_{k}\cos\Theta_{k})\cos\theta_{\vec{k}'\vec{k}}+s\eta i(\cos\Theta_{k}+\cos\Theta_{k})\sin\theta_{\vec{k}'\vec{k}}]\delta\varrho_{k\hat{\vec{k}}',s}^{\bar{\eta}\eta}\notag\\
     &=\frac{ D_n}{8}\frac{k}{ \left\vert v^{0}_{k\eta} \right\vert}[(1+\cos^2\Theta_{k})\delta\varrho_{\vec{k}s}^{\bar{\eta}\eta}+2s\eta i\cos\Theta_{k}\delta\varrho_{\vec{k}s,a}^{\bar{\eta}\eta}]\notag, 
\end{align}
\begin{align}
     \mathcal{ J}^{\bar{\eta}\eta}_{\vec{k}s,\text{off},\text{ra}-2}&=-\frac{\pi \hbar D_n}{V}\frac{V}{(2\pi)^2}\int_{0}^{\infty} k'dk' \int_{0}^{2\pi} d\theta_{\vec{k}'}\delta(\epsilon_{k\eta}-\epsilon_{k'\eta})  \sigma_{\vec{k}\eta,\vec{k}'\eta}^{s}   \sigma_{\vec{k}'\eta,\vec{k}\eta}^{s}  \delta\varrho_{\vec{k}s}^{\bar{\eta}\eta}\\
     &=-\frac{\pi  D_n}{4\pi^2} \frac{k}{ \left\vert v^{0}_{k\eta} \right\vert} \int_{0}^{2\pi} d\theta_{\vec{k}'}\frac{1}{2}\left[1+\cos\Theta_{k}\cos\Theta_{k}\right] \delta\varrho_{\vec{k}s}^{\bar{\eta}\eta} =-\frac{ D_n}{4} \frac{k}{ \left\vert v^{0}_{k\eta} \right\vert} \left(1+\cos^2\Theta_{k}\right) \delta\varrho_{\vec{k}s}^{\bar{\eta}\eta}\notag,
\end{align}
\begin{align}
     \mathcal{ J}^{\bar{\eta}\eta}_{\vec{k}s,\text{off},\text{ra}-3}&=-\frac{\pi \hbar D_n}{V}\frac{V}{(2\pi)^2}\int_{0}^{\infty} k'dk' \int_{0}^{2\pi} d\theta_{\vec{k}'}\delta(\epsilon_{k'\bar{\eta}}-\epsilon_{k\bar{\eta}}) \sigma_{\vec{k}\bar{\eta},\vec{k}'\bar{\eta}}^{s}  \sigma_{\vec{k}'\bar{\eta},\vec{k}\bar{\eta}}^{s}  \delta\varrho_{\vec{k}s}^{\bar{\eta}\eta}\\
     &=-\frac{\pi  D_n}{4\pi^2} \frac{k}{ \left\vert v^{0}_{k\bar{\eta}} \right\vert} \int_{0}^{2\pi} d\theta_{\vec{k}'}\frac{1}{2}\left[1+\cos\Theta_{k}\cos\Theta_{k}\right] \delta\varrho_{\vec{k}s}^{\bar{\eta}\eta} =-\frac{ D_n}{4} \frac{k}{ \left\vert v^{0}_{k\bar{\eta}} \right\vert} \left(1+\cos^2\Theta_{k}\right) \delta\varrho_{\vec{k}s}^{\bar{\eta}\eta} \notag,
\end{align}
\begin{align}
     \mathcal{ J}^{\bar{\eta}\eta}_{\vec{k}s,\text{off},\text{ra}-4}&=\frac{\pi \hbar D_n}{V}\frac{V}{(2\pi)^2}\int_{0}^{\infty} k'dk' \int_{0}^{2\pi} d\theta_{\vec{k}'}\delta(\epsilon_{\vec{k}\bar{\eta}}-\epsilon_{\vec{k}'\bar{\eta}})  \sigma_{\vec{k}\bar{\eta},\vec{k}'\bar{\eta}}^{s}   \sigma_{\vec{k}'\eta,\vec{k}\eta}^{s}  \delta\varrho_{\vec{k}'s}^{\bar{\eta}\eta}\\
     &= \frac{\pi D_n}{4\pi^2}  \frac{k}{ \left\vert v^{0}_{k\bar{\eta}} \right\vert} \int_{0}^{2\pi} d\theta_{\vec{k}'}\frac{1}{2}[(1+\cos\Theta_{k}\cos\Theta_{k})\cos\theta_{\vec{k}'\vec{k}}+s\eta i(\cos\Theta_{k}+\cos\Theta_{k})\sin\theta_{\vec{k}'\vec{k}}]\delta\varrho_{k\hat{\vec{k}'s}}^{\bar{\eta}\eta}\notag\\
     &=\frac{ D_n}{8}\frac{k}{ \left\vert v^{0}_{k\bar{\eta}} \right\vert}[(1+\cos^2\Theta_{k})\delta\varrho_{\vec{k}s}^{\bar{\eta}\eta}+2s\eta i\cos\Theta_{k}\delta\varrho_{\vec{k}s,a}^{\bar{\eta}\eta}]\notag.
\end{align}
\end{widetext}
By substitution of Eqs.~(\ref{fvdvk1}-\ref{fvdvk4}), we attain 
\begin{align}
     \mathcal{ J}^{\bar{\eta}\eta}_{\vec{k}s,\text{off},\text{ra}-1}&+ \mathcal{ J}^{\bar{\eta}\eta}_{\vec{k}s,\text{off},\text{ra}-4}= \frac{ D_nk}{8}(1+\cos^2\Theta_k)\\
     &\times\left(\frac{1}{ \left\vert v^{0}_{k\eta} \right\vert}+\frac{1}{ \left\vert v^{0}_{k\bar{\eta}} \right\vert}\right)\delta\varrho_{\vec{k}s}^{\bar{\eta}\eta}\notag\\
     &+ s\eta i\frac{ D_nk}{4}\cos\Theta_{k} \left(\frac{1}{ \left\vert v^{0}_{k\eta} \right\vert}+\frac{1}{ \left\vert v^{0}_{k\bar{\eta}} \right\vert}\right) \delta\varrho_{\vec{k}s,a}^{\bar{\eta}\eta} .\notag
\end{align}
Besides, we attain 
\begin{align}
    \mathcal{ J}^{\bar{\eta}\eta}_{\vec{k}s,\text{off},\text{ra}-2}+\mathcal{ J}^{\bar{\eta}\eta}_{\vec{k}s,\text{off},\text{ra}-3}&=-\frac{ D_nk}{4}  \left(1+\cos^2\Theta_{k}\right)\\
    &\times \left(\frac{1}{ \left\vert v^{0}_{k\eta} \right\vert}+\frac{1}{ \left\vert v^{0}_{k\bar{\eta}} \right\vert}\right)\delta\varrho_{\vec{ks}}^{\bar{\eta}\eta}.\notag 
\end{align}
Therefore, we attain the intraband collision term arising from the single-band scattering  as following:
\begin{align}
     \mathcal{ J}^{\bar{\eta}\eta}_{\vec{k}s,\text{off},\text{ra}}= -\frac{1}{\hbar}\Gamma_{k} \delta\varrho^{\bar{\eta}\eta}_{\vec{k}s}+is\eta\frac{1}{\hbar}\Gamma^a_{k}\delta\varrho^{\bar{\eta}\eta}_{\vec{k}s,a}, 
\end{align}
where the normal and anomalous scattering rate are given by
\begin{align}
     \Gamma_{k}&=\frac{\hbar D_nk}{8}  \left(1+\cos^2\Theta_{k}\right)\left(\frac{1}{ \left\vert v^{0}_{k\eta} \right\vert}+\frac{1}{ \left\vert v^{0}_{k\bar{\eta}} \right\vert}\right)\\
     &=\frac{1}{8}\left(1+\cos^2\Theta_{k}\right)\left(\frac{\hbar}{\tau^{0}_{k\eta}}+\frac{\hbar}{\tau^{0}_{k\bar{\eta}}}\right)\notag, 
\end{align}
\begin{align}
     \Gamma^a_{k}&=\frac{\hbar D_nk}{4} \cos\Theta_{k} \left(\frac{1}{ \left\vert v^{0}_{k\eta} \right\vert}+\frac{1}{ \left\vert v^{0}_{k\bar{\eta}} \right\vert}\right)\\
     &=\frac{1}{4}\cos\Theta_{k}\left(\frac{\hbar}{\tau^{0}_{k\eta}}+\frac{\hbar}{\tau^{0}_{k\bar{\eta}}}\right)\notag. 
\end{align}

\end{document}